\documentclass[aps,pre,twocolumn,groupedaddress]{revtex4-1}

\usepackage{graphicx}
\usepackage{amsmath,amssymb}

\begin{document}

\title{WKB-type-of approximation for rare event statistics in reacting systems}

\author{Andreas M{\"u}hlbacher}
\email{andreas.muehlbacher@uni-due.de}

\author{Thomas Guhr}
\email{thomas.guhr@uni-due.de}
\affiliation{Faculty of Physics, University of Duisburg-Essen, Lotharstr. 1, 47048 Duisburg, Germany}

\date{\today}

\begin{abstract}
We calculate the probabilities to find systems of reacting particles in states which largely deviate from typical behavior. The rare event statistics is obtained from the master equation which describes the dynamics of the probability distribution of the particle number. We transform the master equation by means of a generating function into a time-dependent ``Schr\"odinger equation''. Its solution is provided by a separation ansatz and an approximation for the stationary part which is of Wentzel-Kramers-Brillouin (WKB) type employing a small parameter. The solutions of the ``classical'' equations of motions and a saddle point approximation yield the proper generating function. Our approach extends a method put forward by Elgart and Kamenev. We calculate the rare event statistics for systems where the dynamics cannot be entirely analyzed in an analytical manner. We consider different examples.
\end{abstract}

\maketitle

\section{Introduction}
Rare events are important in many situations, especially when their consequences are extreme \cite{PubRev2018}. Often, demographic noise or intrinsic fluctuations which are based on stochasticity cause such events. There is a wide variety of applications where stochasticity plays a key role, such as chemical reactions \cite{cardelli2016,van2007}, population dynamics \cite{bartlett1960,allen2010,nisbet2003,naasell2011,krug2011}, epidemiology \cite{andersson2012,allen2008,keeling2008,black2010,allen2015} and financial markets \cite{muehlbacher2018,gardiner2009} to name a few examples. Another example for stochasticity is non-demographic, i.e. extrinsic, noise which arises due to interactions between the considered system and a noisy environment \cite{mao2002,elijah2015,lande1993,kamenev2008}. If one ignores the noise, many of such systems can be described by rate equations to obtain a macroscopic, deterministic description of average quantities of the system, particularly mean concentrations. This is referred to as the mean field approximation. However, we are interested in large deviations from a typical system behavior. For example, in population dynamics, this is the case when the average size of the population is large and we are interested in low population numbers. The state of this system is given by the population number at a distinct time. The probabilistic description of a stochastic system in such a state is given by the master equation \cite{gillespie1992,weber2017}. Despite its simplicity it is not possible, apart from a few exceptions \cite{jahnke2007,mcquarrie1963,mccquarrie1964,laurenzi2000}, to solve the master equation analytically \cite{walczak2012,jenkinson2012,schnoerr2017}. Hence, other methods are called for. There are many kinds of numerical simulations \cite{li2008,pahle2009,gillespie2013,Mauch2011} to model the stochastic kinetics. The stochastic simulation algorithm \cite{gillespie1976,gillespie1977,gillespie2007} simulates sample paths of the underlying stochastic process, especially for chemical reactions. Due to limited computational resources simulations are not appropriate for large systems and not for the determination of rare events either. An approximation of the master equation is obtained by a Fokker-Planck equation \cite{risken1996}. To this end one has to apply a van Kampen system size expansion. The Fokker-Planck equation is reliable only for small deviations from the mean field approximation. It fails to give an accurate description of large deviations from the typical evolution of a system \cite{doering2005,gaveau1996}. Elgart and Kamenev \cite{ElgartKamenev2004} introduced an asymptotic method to calculate the rare event statistics in reaction-diffusion systems by formally relating it to semiclassics. A quantum problem can sometimes be solved by expanding to lowest order in $\hbar$ around the corresponding classical equations of motion. A ``Hamiltonian'' formulation of reaction-diffusion systems is developed which reformulates the master equation by means of a generating function as a ``Schr\"odinger equation''. The ``semiclassical'' dynamics of the corresponding Hamiltonian provides all the information necessary for the analysis of rare event statistics. The existence of a small parameter allows the treatment analogous to a Wenzel-Kramers-Brillouin (WKB) approximation \cite{landau1977,miller1953,messiah1964,griffiths2018,bender2013}. The WKB approximation is applied to the evolution equation of the generating function. For related studies on the WKB approximation see \cite{assaf2006,assaf2010,assaf2017,beer2015,beer2016,dykman1994,gang1987,kubo1973,peters1989} and for reviews see \cite{assaf2017rev,ovaskainen2010,bressloff2017}. For stochastic population models the WKB approximation allows one also to calculate the mean extinction times and probabilities \cite{assaf2007ex,assaf2010ext,kessler2007,assaf2009ex} and for switching rates in multistep reactions see \cite{escudero2009}. 

Here, we consider single-species chemical reactions which can be described by master equations that give the time evolution of the probability to find a distinct number of particles at a given time. One way of dealing with such systems is a spectral formulation and a stationary WKB approximation, see \cite{assaf2006}. A different method is the time-dependent semiclassical approximation \cite{ElgartKamenev2004}. Both methods have different regimes of validity and accuracy. The time-dependent semiclassical approximation is accurate for $1\ll n\leq\langle n\rangle$ which also means for not too long times. In the region $n>\langle n\rangle$ the accuracy of this method breaks down and the spectral formulation and stationary WKB approximation is better suited. We generalize the time-dependent model \cite{ElgartKamenev2004} in such a way that reactions with more than two particles of one species can be analyzed. This is not possible in the original method since the corresponding equations cannot be solved analytically. Moreover, we include the calculation of the pre-exponential factor of the distribution. This was disregarded in the original method. Our main focus lies on the probability of rare events, i.e., to find our system in a state far away from the typical behavior. By means of the probabilities we are able to estimate many quantities such as the average extinction time and the lifetime distribution. Even though large deviations from a typical system behavior may be hardly observable, their probabilities are interesting for anyone who has to compensate probable risks which come along with these rare events. In more detail we consider systems which can have, depending on the initial condition, an absorbing state with zero particles left. In general, results for such systems are unavailable in analytical form. However, with our approach, we are able to compute the solution partially in analytical and partially in numerical form. Depending on the reaction scheme, we have to scale the corresponding parameters of the master equation in order to perform the WKB approximation in a proper way. We consider different types of reactions where we combine death or single annihilation, binary annihilation and triple annihilation. As a result we find a very good accordance between the WKB approximation and the exact solution of the master equation especially in the tail of the distribution. 

The paper is structured as follows. In section \ref{sec theorydeeper} we go into the details of the approach and generalize it in such a way that more complex reactions like higher order annihilations can be analyzed. In section \ref{sec examples} we apply the method on a set of examples which can either be solved exactly or by means of the WKB approximation and numerical calculus. We conclude in section \ref{sec conclude}.

\section{Deeper look at the method}\label{sec theorydeeper}
Consider a system consisting of identical particles which can react with each other according to different reaction schemes $i=1,\dotsc, m$. The chemical reactions can be described by master equations which give the time evolution of the probability to find a given number of particles in the system at a given time. A particular reaction occurs with probability $\lambda_i \Delta t$ with $\lambda_i\ll 1$ in the time interval $[t,t+\Delta t]$, where $\lambda_i$ is the specific probability rate constant. Importantly, all $\lambda_i$ are independent of $\Delta t$ or the considered time interval. Since each particle may react with each other the system is fully described by the following master equation
\begin{equation}
\frac{d}{dt} P_n(t)=\sum^{m}_{i=1}\left(\lambda_ih_i(n-\nu_i)P_{n-\nu_i}(t)-\lambda_ih_i(n)P_n(t)\right)
\label{eq:mastereqgen}
\end{equation}
where $P_n(t)$ denotes the probability to find $n$ particles at time $t$, $h_i(n)$ is the number of combinations of reacting particles in the system under reaction scheme $i$ when $n$ particles are in the system and $\nu_i$ is the change of particle number when the reaction $i$ occurs, see \cite{gillespie1992}. In order to obtain an unique solution of the master equation we have to specify an initial condition. This can be any kind of normalized distribution like the Poisson distribution or for our sake we use a fixed particle number $n_0$ and hence $P_n(0)=\delta_{n,n_0}$, with the Kronecker delta $\delta_{n,n_0}$.

We introduce the auxiliary variable $\xi$ which formally plays the r\^{o}le as a ``position'' and the generating function
\begin{equation}
 G(\xi,t)=\sum^{\infty}_{n=0}\xi^nP_n(t)\;.
\label{eq genfunc}
\end{equation}
The generating function has to fulfill the initial condition $P_n(0)$ which gives $G(\xi,0)=\xi^n$. Furthermore, from the conservation of probability, we find $G(1,t)=1$. By means of the generating function we are able to calculate the average particle number
\begin{equation}
 \langle n\rangle=\sum^{\infty}_{n=0}nP_n(t)=\left.\frac{\partial}{\partial \xi}G(\xi,t)\right|_{\xi=1}
 \label{eq partnumber}
\end{equation}
and the probability
\begin{equation}
 P_n(t)=\frac{1}{n!}\left.\frac{\partial^n}{\partial\xi^n}G(\xi,t)\right|_{\xi=0}.
 \label{eq probderiv}
\end{equation}
If the particle number $n$ is large it is convenient to use Cauchy's integral formula
\begin{equation}
 P_n(t)=\frac{1}{2\pi i}\oint d\xi\, G(\xi,t)\xi^{-n-1}
 \label{eq cauchy}
\end{equation}
where the integration is performed over a closed contour encircling $\xi=0$. Multiplying both sides of \eqref{eq:mastereqgen} with $\xi^n$ and summing over all $n$ yields the partial differential equation
\begin{equation}
 \frac{\partial}{\partial t}G(\xi,t)=\hat{\mathcal{L}}G(\xi,t)\;,
 \label{eq partdiffLgenfunc}
\end{equation}
where $\hat{\mathcal{L}}$ is a linear differential operator that includes powers of $\partial/\partial \xi$. The requirement of analyticity of $G(\xi,t)$ yields ``self-generated'' boundary conditions. Besides the universal boundary condition $G(1,t)=1$ the others are specific to the problem at hand. However, there is always the physical initial condition which is determined by the value of $P_n(0)$ and hence it is $G(\xi,0)$ that needs to be considered. The partial differential equation \eqref{eq partdiffLgenfunc} can formally be written as a time-dependent ``Schr\"odinger equation'' with imaginary time
\begin{equation}
 i \lambda\frac{\partial}{\partial it}G(\xi,t)=\hat{H}G(\xi,t)\;,
 \label{eq schroedinger}
\end{equation}
where the right hand side is the ``Hamilton operator'' $\hat{H}$ and we define $\lambda=\lambda_k$ for a fixed $k$. Now, we employ the formal analogy of the probability $\lambda$ with Planck's constant $\hbar$. Moreover, in analogy to quantum mechanics we define the momentum operator 
\begin{equation}
 \hat{\pi}=-i\lambda \frac{\partial}{\partial \xi}\;,
 \label{eq counteroperator}
\end{equation}
which will acquire the meaning of a counting operator. By inserting the ansatz $G(\xi,t)=\varphi(\xi)\psi(it)$ into \eqref{eq schroedinger} we can separate the variables $\xi$ and $t$ and find the two equations
\begin{eqnarray}
  i \lambda\frac{\partial}{\partial it}\psi(it)&&=E\psi(it) \label{eq separat time}\\ 
	\hat{H}\varphi(\xi)&&=E\varphi(\xi).\label{eq separat stat}
\end{eqnarray}
We will interpret the constant $E$ as energy or Hamilton function of our system. Up to this point we consider a probabilistic description of our problem, i.e., both variables $\xi$ and $t$ can be chosen independently from each other. However, the separation ansatz leads to the formal problem $G(1,t)=\varphi(1)\psi(it)=1$ where $t$ cannot be chosen arbitrarily. We will solve this problem by a semiclassical approximation in which we use the classical equations of motion that give an explicit dependence of both variables $\xi$ and $t$. The solution of Eq.~\eqref{eq separat time} is
\begin{equation}
 \psi(it)=\psi(0)e^{Et/\lambda}\;.
\label{eq soltimebinary}
\end{equation}
We solve the stationary ``Schr\"odinger equation'' \eqref{eq separat stat} with a WKB approximation. We insert the ansatz
\begin{equation}
\varphi(\xi)=A(\xi)\exp(iS(\xi)/\lambda)
\label{eq wkbansatz}
\end{equation}
and separate the resulting equation into its real and imaginary parts. To solve the pair of differential equations we use the standard WKB assumption that all terms of second order or higher in the small parameter, here $\lambda$, can be neglected. The WKB approximation requires that the ``quantum'' fluctuations are weak, which is true as long as $\langle n(t)\rangle\gg 1$, i.e., for times not too long. In this regime we can apply the condition $\lambda\ll 1$. By putting all solutions of the differential equations together we find the solution for the generating function $G(\xi,t)$. This solution depends on a constant which has to be determined by the initial condition $G(\xi,0)$. Moreover, the generating function also depends on the energy $E$ of our system. To make progress we have to determine the value of $E$.

Up to this point we used a probabilistic interpretation of the system where the variables $\xi$ and $t$ can be chosen independently from each other. To approximate the function $G(\xi,t)$ and determine the value of $E$ we now go into semiclassics. The energy is given by the Hamilton function $H$ which can be inferred from the ``Hamilton operator'' $\hat{H}$ and reads
\begin{equation}
 H=H(\xi,\pi)=E
 \label{eq energy}
\end{equation}
where $\pi$ is the classical ``momentum''. The energy is an integral of motion with $dE/dt=0$. The classical equations of motion in imaginary time are the ``Hamilton equations''
\begin{eqnarray}
 \frac{d\xi}{d it}&&=\frac{\partial H}{\partial \pi}\label{eq classgenxi}\\
 \frac{d\pi}{d it}&&=-\frac{\partial H}{\partial \xi}\;.
 \label{eq classgenpi}
\end{eqnarray}
Due to the classical equations of motion \eqref{eq classgenxi} and \eqref{eq classgenpi}, the variables $\xi$ and $t$ are no longer independent of each other. From this point, $\xi$ and $\pi$ are viewed as functions of $t$. We solve the energy $E$ for $\pi$ and insert the result into \eqref{eq classgenxi}. Now, the first equation of motion contains only $\xi$ and the constant energy $E$ and we find
\begin{equation}
\frac{d\xi}{d it}=\tau(\xi,E)\;,
\label{eq eqmotgener}
\end{equation}
which gives the explicit classical dependence $\xi=\xi(t)$. Importantly, for the determination of the initial condition of the generating function we find $\xi(0)=\xi_0$. In the original method \cite{ElgartKamenev2004} the assumption $\xi_0=1$ was made. In section \ref{sec one} we show in a direct comparison between our and the original method the benefit of taking the variable $\xi_0$ into account, rather than making the approximation $\xi_0=1$. We do not solve the second equation of motion \eqref{eq classgenpi}, instead we consider the mean field dynamics. If we are interested in the average particle number $\langle n\rangle$, we have to know the generating function in the vicinity of $\xi=1$, see Eq.~\eqref{eq partnumber}. This is the constant mean field solution $\bar\xi=1$ which solves Eq.~\eqref{eq classgenxi} because every permissible Hamilton function must satisfy the condition $H(1,\pi)=0$ due to the normalization of probability. By application of the mean field solution the second equation of motion \eqref{eq classgenpi} has the solution $\bar\pi(t)$ where we specify the initial condition $\bar\pi(0)=\pi_0$. Acting with the previously defined momentum operator $\hat{\pi}$ on the generating function at the mean field solution gives
\begin{equation}
 \hat{\pi} G(\xi,t)\Big|_{\xi=1}=\frac{\lambda}{i}\left.\frac{\partial G(\xi,t)}{\partial \xi}\right|_{\xi=1}=\frac{\lambda}{i}\langle n\rangle=\bar{\pi}(t)\;.
\end{equation}
The last equality holds because in the mean field approximation quantum mechanics has to coincide with classical mechanics. Thus, we call the operator $\hat{\pi}$ counting operator which gives the average particle number for a given time. As $\pi_0$ has to be constant for all times we can specify its value at time $t=0$ where, due to the initial condition, we find $\langle n\rangle_{t=0}=n_0$ and therefore
\begin{equation}
 \pi_0=\frac{\lambda}{i}n_0\;.
\end{equation}
We note that $E$ is constant and can also be expressed in terms of $\xi(0)=\xi_0$ and $\pi(0)=\pi_0=-i\lambda n_0$, i.e., $E=E(\xi_0)=E_0$. Eq.~\eqref{eq eqmotgener} can be solved by integration. If we allow for complex reactions the function $\tau(\xi,E)$ will become also complex. In many cases it might not be possible to solve the differential equation \eqref{eq eqmotgener} analytically. Instead a numerical solution has to be taken into account. Examples therefore are the reaction which combines binary and single annihilation and the combined reaction up to third order annihilation which are discussed in section \ref{sec: bin and single} and \ref{sec: third order}, respectively. This issue is the starting point of our more general approach to calculate the event statistics.

The formal solution of Eq.~\eqref{eq eqmotgener} is
\begin{equation}
\int^{\xi}_{\xi_0}{\frac{d\xi'}{\tau(\xi',E)}}=it\;.
\label{eq class1formal}
\end{equation}
The initial condition $G(\xi,0)=\xi^{n_0}_0$ enables us to determine the constant in the generating function $G(\xi,t)$. Before we determine the value $\xi_0$ we first calculate the probability $P_n(t)$ by means of a saddle point approximation. At this point we have to make an important remark. As already discussed, the starting point of our analysis is a probabilistic description of the process. We proceed by using semiclassical methods in order to find an approximation for the generating function. Therefore, we derive equations of motion that describe the classical trajectories for a constant energy $E$. The classical trajectories give an explicit dependence of the underlying variables which in our case are $\xi$, $\pi$ and $t$. They cannot be chosen independently from each other. The value of the energy determines which trajectory in the phase space describes the relationship between $\xi$ and $t$, i.e., $\xi=\xi(t)$. Eventually, we are interested in probabilities $P_n(t)$ again. Therefore, we have to leave the semiclassical description behind and turn back to a probabilistic description. In a probabilistic description the variables $\xi$ and $t$ are independent of each other and can also be chosen independently. Classically this means that we do not move along the trajectories anymore. This only applies if the energy depends on both variables, i.e., $E=E(\xi,t)$. In other words, when we go back from the semiclassical into the probabilistic description the energy does not remain constant. In that sense it is more intuitive to name the quantity $E$ Hamilton function instead of energy. This non-constant Hamilton function can be explained by a short example. In a conservative system the energy is fully determined by the initial conditions. It however differs for different initial conditions. In that sense, the Hamilton function describes, at the same time, a constant of motion and the same system at different energies. Hence, derivatives $E'=\partial E/\partial \xi$ have to be taken into account when the saddle point approximation is performed.

The contour integral can be written as
\begin{equation}
P_n(t)=\frac{1}{2\pi i}\oint{d\xi\, g(\xi)\exp(f(\xi,E))}
\end{equation}
with the function $g(\xi)$ and the ``free energy'' $f(\xi,E)$ that are determined by the generating function $G(\xi,t)$ and the factor $\xi^{-n-1}$ in Cauchy's integral formula \eqref{eq cauchy}. We add the term $\xi^{-n}$ into the exponential because of its factor $n$ which is supposed to be large. We calculate the integral by means of a saddle point approximation. This approximation is justified because we have a small $\lambda$ and assume large $n$. Furthermore, we are interested in times $t$ where $\langle n(t)\rangle$ is sufficiently smaller than $n_0$ but $\langle n(t)\rangle\gg 1$ still holds.

The saddle point approximation requires $f'(\xi_s,E)=0$ and we find with a non-constant energy $E(\xi,t)$
\begin{equation}
f'(\xi,E)=\omega(\xi,E,E')\;.
\end{equation}
For convenience we drop the arguments of $E(\xi,t)$. Pretending that the analytical solution of \eqref{eq eqmotgener} is not known, i.e., when we have to evaluate the integral in \eqref{eq class1formal} numerically, we have to determine $E'$ in order to solve the saddle point equation $f'(\xi_s,E)=0$. We obtain the derivative of $E$ by deriving the solution of the classical equation of motion \eqref{eq class1formal} with respect to $\xi$ under the condition that the energy depends on $\xi$ and the time $t$ does not
\begin{equation}
0=\frac{\partial}{\partial \xi}\int^{\xi}_{\xi_0}{\frac{d\xi'}{\tau(\xi',E(\xi,t))}}\;.
\end{equation}
We remark that the integration is performed over the variable $\xi'$ which is due to semiclassics not included in $E(\xi,t)$. By using the Leibniz rule we find
\begin{eqnarray}
E'&&=\frac{\partial E(\xi,t)}{\partial \xi}\nonumber\\
&&=\left(\tau(\xi,E(\xi,t))\int^{\xi}_{\xi_0}{\frac{d\xi'}{\tau^2(\xi',E(\xi,t))}\frac{\partial}{\partial E}\tau(\xi',E(\xi,t))}\right)^{\!\!-1}\nonumber\\
&&=\kappa(\xi,E)\;,
\label{eq eprime}
\end{eqnarray}
which again is a function of $\xi$ and $E$. The energy $E$ can be expressed in dependence of $\xi_0$, i.e., $E=E(\xi_0)$. Due to this relation $E'$ can also be expressed in terms of $\xi$ and $\xi_0$, i.e., $E'=\kappa(\xi,\xi_0)$. Now, we can reduce the derivative $f'(\xi,E)=\omega(\xi,\xi_0)$ on the two variables $\xi$ and $\xi_0$ and the saddle point condition becomes
\begin{equation}
\omega(\xi_s,\xi_0)=0\;.
\label{eq depxi0xis}
\end{equation}
From its solution we obtain the relationship between $\xi_s$ and $\xi_0$, i.e., $\xi_0=\xi_0(\xi_s)$. Together with Eq.~\eqref{eq class1formal} this allows us to calculate a numeric value for $\xi_s$, by solving the integral equation
\begin{equation}
\int^{\xi_s}_{\xi_0(\xi_s)}{\frac{d\xi'}{\tau(\xi',E_0(\xi_0(\xi_s)))}}=it\;.
\label{eq solinteqgeneral}
\end{equation}
We note that the integral stems from the semiclassical part of our analysis, hence the energy $E$ has to remain constant during integration. Furthermore we changed the upper limit in \eqref{eq class1formal} to $\xi_s$ which is valid, as the constant value of the energy depending on $\xi_s$ has to be chosen accordingly. Finding a solution for $\xi_s$ can be demanding, depending on the function $\tau(\xi,E)$ and the solution $\xi_0(\xi_s)$. Once, a solution for $\xi_s$ is found the value of $\xi_0$ can be calculated. The last piece missing for the saddle point approximation is the second derivative of the free energy
\begin{eqnarray}
f''(\xi,E)=&&\frac{\partial \omega(\xi,E,E')}{\partial \xi}+\frac{\partial \omega(\xi,E,E')}{\partial E}E'\nonumber\\
&&+\frac{\partial \omega(\xi,E,E')}{\partial E'}E''\;,
\end{eqnarray}
with the second derivative of the energy
\begin{widetext}
\begin{eqnarray}
E''=\frac{\partial^2 E(\xi)}{\partial \xi^2}=-(E')^2&&\left\{\frac{1}{E'\tau(\xi,E(\xi))}\frac{\partial}{\partial \xi}\tau(\xi,E(\xi))+\frac{1}{\tau(\xi,E(\xi))}\frac{\partial}{\partial E}\tau(\xi,E(\xi))\right.\nonumber\\
&&\left.+E'\tau(\xi,E(\xi))\int\limits^{\xi}_{\xi_0}{\frac{d\xi'}{\tau^2(\xi',E(\xi))}\left[\frac{\partial^2}{\partial E^2}\tau(\xi',E(\xi))-\frac{2}{\tau(\xi',E(\xi))}\left(\frac{\partial}{\partial E}\tau(\xi',E(\xi))\right)^2\right]}\right\}\;.
\end{eqnarray}
\end{widetext}
The value of $E$ is determined by $E(\xi_0)=E_0$. Combining all pieces yields the final result for the probability
\begin{eqnarray}
P_n(t)&&=\frac{1}{\sqrt{2\pi}}\frac{g(\xi_s)}{\sqrt{\left|f''(\xi_s,E)\right|}}\exp\left(f(\xi_s,E_0)\right)\:,
\end{eqnarray}
if $f''(\xi,E)$ is always real. This formalism works for every kind of reaction scheme. However, if the reactions become too complex, e.g. by involving higher order annihilations, it might become impossible to solve the required equations due to technical complications.

\section{Some examples}\label{sec examples}

\subsection{Binary annihilation revisited}\label{sec one}
For the convenience of the reader we show how the known results of binary annihilation, studied in \cite{ElgartKamenev2004}, fit into our generalized approach. This model can be solved analytically and exactly \cite{mccquarrie1964}.

\begin{figure}[h]
 %\centering
 \includegraphics[width=0.48\textwidth]{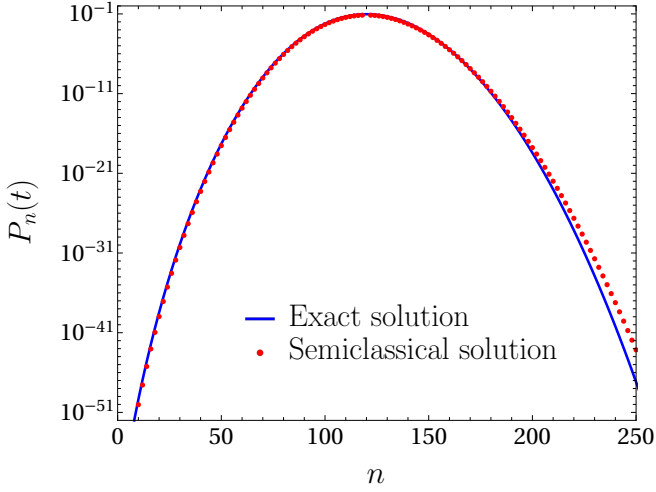}
  \caption{\label{fig compbinary}Probability density distribution for the binary annihilation, see Eq.~\eqref{eq master}, at $t=0.5$ on a logarithmic scale. Parameters are chosen as $n_0=300$ and $\lambda=0.01$.}
\end{figure}

\begin{figure}[t]
 %\centering
 \includegraphics[width=0.48\textwidth]{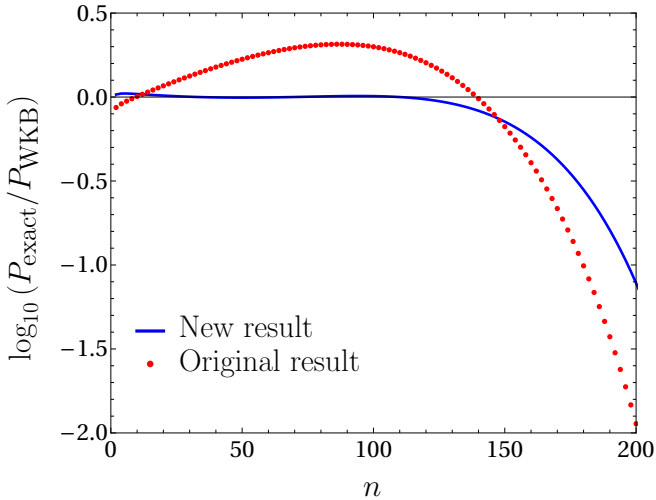}
  \caption{Decadic logarithm of the ratio of the exact solution and the time-dependent WKB approximations. The solid line shows the new approximation \eqref{eq binapproxP}, the circles show the original result. Parameters are chosen as $n_0=300$, $t=1$ and $\lambda=0.01$.}
  \label{fig compNewOrig}
\end{figure}

We consider a system where only the binary annihilation can possibly occur with probability rate $\lambda$. Once, this reaction takes place two particles form an inert aggregate. Since each particle may react with each other the system is fully described by the following master equation
\begin{equation}
 \frac{d}{dt} P_n(t)=\frac{\lambda}{2}\left((n+2)(n+1)P_{n+2}(t)-n(n-1)P_n(t)\right)\;,
 \label{eq master}
\end{equation}
where $P_n(t)$ denotes the probability to find $n$ particles at time $t$. According to Eq.~\eqref{eq schroedinger} the corresponding time-dependent ``Schr\"odinger equation'' with imaginary time is
\begin{equation}
 i \lambda\frac{\partial}{\partial it}G(\xi,t)=\frac{\lambda^2}{2}(1-\xi^2)\frac{\partial^2}{\partial \xi^2}G(\xi,t)\;,
% \label{eq schroedinger}
\end{equation}
where the right hand side can also be written in terms of the ``Hamilton operator''
\begin{equation}
 \hat{H}=\frac{\lambda^2}{2}(1-\xi^2)\frac{\partial^2}{\partial \xi^2}=-\frac{1}{2}(1-\xi^2)\hat{\pi}^2\;.
\end{equation}
The separation ansatz $G(\xi,t)=\varphi(\xi)\psi(it)$ yields equation \eqref{eq soltimebinary} for the time-dependent part and 
\begin{equation}
	E\varphi(\xi)=\frac{\lambda^2}{2}(1-\xi^2)\frac{\partial^2}{\partial \xi^2}\varphi(\xi)
\end{equation}
for the stationary part. We solve the stationary ``Schr\"odinger equation'' with a WKB approximation. We insert the ansatz \eqref{eq wkbansatz} and separate the resulting equation into its real and imaginary parts which yields
\begin{eqnarray}
EA(\xi)&&= -\frac{\lambda^2}{2}(\xi^2-1)\left(A''(\xi)-\frac{1}{\lambda^2}A(\xi)\left(S'(\xi)\right)^2\right)\label{eq wkb1}\\
 0&&=2A'(\xi)S'(\xi)+A(\xi)S''(\xi)\;.
 \label{eq wkb2}
\end{eqnarray}
Eq.~\eqref{eq wkb2} can be simplified
\begin{equation}
 \frac{d}{d\xi}A^2(\xi)S'(\xi)=0
\end{equation}
which yields
\begin{equation}
 A(\xi)=\frac{\tilde{c}}{\sqrt{S'(\xi)}}
\end{equation}
with a constant $\tilde{c}$. In Eq.~\eqref{eq wkb1} we use the standard WKB assumption that all terms of second order in the small parameter, here $\lambda$, can be neglected. This gives the simplified differential equation
\begin{equation}
 \frac12 (\xi^2-1)\left(S'(\xi)\right)^2=E\;.
\end{equation}
Its solution is
\begin{equation}
 S(\xi)-S(\xi_0)=i\sqrt{2E}(\arccos\xi-\arccos\xi_0)\;,
\label{eq solSwkb}
\end{equation}
with some initial value $\xi_0$. Now, we can put all terms together and obtain the generating function
\begin{eqnarray}
 G(\xi,t)=&&c\left(\frac{1-\xi^2}{2E}\right)^{1/4}\nonumber\\
&&\times\exp\!\left(\!\!-\frac{\sqrt{2E}}{\lambda}(\arccos\xi-\arccos\xi_0)+\!\frac{Et}{\lambda}\right),
 \label{eq genfunccE}
\end{eqnarray}
where all constants have been put into $c$. Before determining this constant by the initial condition of the generating function, we first go into semiclassics. The Hamilton function is given by 
\begin{equation}
 H=E=\frac12(\xi^2-1)\pi^2\;.
 \label{eq energybin}
\end{equation}
The classical equations of motion in imaginary time are
\begin{eqnarray}
 \frac{d\xi}{d it}&&=\frac{\partial H}{\partial \pi}=(\xi^2-1)\pi\label{eq class1}\\
 \frac{d\pi}{d it}&&=-\frac{\partial H}{\partial \xi}=-\xi \pi^2\;.
 \label{eq class2}
\end{eqnarray}
We solve Eq.~\eqref{eq energybin} for $\pi$ and insert the solution into Eq.~\eqref{eq class1}. The ensuing differential equation
\begin{equation}
 \frac{d\xi}{d it}=\sqrt{2E(\xi^2-1)}
\label{eq diffbinannex}
\end{equation}
has the solution
\begin{equation}
 \arccos\xi-\arccos\xi_0=\sqrt{2E}t\;,
 \label{eq solclass1}
\end{equation}
which is analogous to Eq.~\eqref{eq class1formal}. The mean field dynamics gives us the average particle number. By application of the mean field solution $\bar\xi=1$ the second equation of motion \eqref{eq class2} has the solution
\begin{equation}
 \bar\pi(t)=\frac{\lambda}{i}\langle n\rangle=\frac{\pi_0}{it\pi_0+1}\;.
\end{equation}
The average particle number is
\begin{equation}
 \langle n(t)\rangle=\frac{n_0}{n_0\lambda t+1}\approx \frac{1}{\lambda t}
\label{eq averagepartnumber}
\end{equation}
for $n_0\gg 1$.
The energy can be written as
\begin{equation}
E=E_0=\frac12 (\xi^2_0-1)\pi^2_0=\frac{\lambda^2}{2}n^2_0(1-\xi^2_0)
\label{eq energyconstant}
\end{equation}
Now, we are able to determine the constant $c$ in Eq.~\eqref{eq genfunccE} with the initial condition $G(\xi,0)=\xi^n_0$
\begin{equation}
c=\xi^{n_0}_{0}\left(\frac{2E}{1-\xi^2_0}\right)^{1/4}\;.
\end{equation}
Obviously, the normalization $G(1,t)=1$ remains preserved. We calculate the probability $P_n(t)$ by means of the contour integral \eqref{eq cauchy} and a saddle point approximation. The contour integral now reads
\begin{equation}
P_n(t)=\frac{1}{2\pi i}\frac{\xi^{n_0}_0}{(1-\xi^2_0)^{1/4}}\oint{d\xi\, g(\xi)\exp(f(\xi,E))}
\end{equation}
with
\begin{eqnarray}
g(\xi)&&=\frac{1}{\xi}(1-\xi^2)^{1/4}\\
f(\xi,E)&&=-\frac{\sqrt{2E}}{\lambda}(\arccos\xi-\arccos\xi_0)+\frac{Et}{\lambda}-n\ln\xi.
\end{eqnarray}
The saddle point approximation requires $f'(\xi_s,E)=0$ and we find with a non-constant Hamilton function $E(\xi,t)$
\begin{eqnarray}
f'(\xi,E)=&&-\frac{E'}{\lambda}\left(\frac{\arccos\xi-\arccos\xi_0}{\sqrt{2E}}-t\right)\nonumber\\
&&+\frac{\sqrt{2E}}{\lambda}\frac{1}{\sqrt{1-\xi^2}}-\frac{n}{\xi}\;.
\end{eqnarray}
For convenience we drop the arguments of $E(\xi,t)$. We see that inserting the solution \eqref{eq solclass1} of the classical equation of motion deletes the first bracket and yields
\begin{equation}
 \frac{1}{\lambda t}(\arccos\xi_s-\arccos\xi_0)\frac{1}{\sqrt{1-\xi^2_s}}=\frac{n}{\xi_s}\;.
 \label{eq solsaddle}
\end{equation}
Interestingly, the $E'$ term has been removed and with Eq.~\eqref{eq energyconstant} we immediately obtain the relation $\xi_0=\xi_0(\xi_s)$ according to Eq.~\eqref{eq depxi0xis}
\begin{equation}
 \xi_0=\sqrt{1-\frac{n^2}{n^2_0}\frac{1-\xi^2_s}{\xi^2_s}}\;.
 \label{eq x0xs}
\end{equation}
Combining Eqs.~\eqref{eq x0xs}, \eqref{eq energyconstant} and \eqref{eq solclass1} yields an implicit equation for $\xi_s$ which has to be solved by numerical methods. Having found the solution for $\xi_s$, a value for $\xi_0$ can be obtained by Eq.~\eqref{eq x0xs}. In order to obtain the second derivative of the free energy we need $E'$ which is according to Eq.~\eqref{eq eprime}
\begin{equation}
E'=-\sqrt{2E}/(t\sqrt{1-\xi^2})\;.
\end{equation}
This result can directly be obtained by deriving Eq.~\eqref{eq solclass1}. The second derivative of $f(\xi)$ at $\xi_s$ can be simplified to
\begin{eqnarray}
 f''(\xi_s)=\frac{n\lambda t-\xi^2_s}{\lambda t(1-\xi^2_s)\xi^2_s}\approx \frac{n-\langle n\rangle\xi^2_s}{(1-\xi^2_s)\xi^2_s}\;,
\end{eqnarray}
which is always real for real $\xi_s$. Combining all pieces yields the final result for the probability
\begin{eqnarray}
 P_n(t)=&&\xi^{n_0}_0\sqrt{\frac{n_0\xi_s}{2\pi n}}\left|\frac{1-\xi^2_s}{n-\langle n\rangle\xi^2_s}\right|^{1/2}\nonumber\\
&&\times\exp\left(-\frac{n^2}{2}\lambda t\frac{1-\xi^2_s}{\xi^2_s}-n\ln\xi_s\right)\;.
\label{eq binapproxP}
\end{eqnarray}
In Fig.~\ref{fig compbinary} we compare the approximation \eqref{eq binapproxP} with the exact solution. We set $n_0=300$, $\lambda=0.01$ and $t=0.5$. The exact solution is calculated by numerical integration of the master equation. The maximum probability is around the average particle number which according to Eq.~\eqref{eq averagepartnumber} is $\langle n(0.5)\rangle=120$. Overall the approximation coincides over many orders of magnitude very well with the exact solution.
We compare the new result \eqref{eq binapproxP} with the original one. In fact, we do not use the original result of \cite{ElgartKamenev2004}, where the prefactor was included manually, we use the time-dependent solution of \cite{assaf2006} that includes the prefactor by calculation. Fig.~\ref{fig compNewOrig} shows on a log-scale the ratios of the exact result and the semiclassical approximation for both, the new result \eqref{eq binapproxP} and the original result. Clearly, in the region $n\leq\langle n\rangle$ we see the high accuracy of the new result which outperforms the original result. For large $n$ the accuracy of the time-dependent semiclassical approximation deteriorates. 

\subsection{Single annihilation}
The simplest reaction is the single annihilation, where one particle forms an inert aggregate with probability rate $\lambda$. In terms of population dynamics this reaction is called death process. This process can be solved analytically and it is well known in literature, see, e.g.~\cite{chang1968,jahnke2007}. Nevertheless, we apply the method to give a different representation how the model can be solved. Its master equation reads
\begin{equation}
\frac{d}{dt}P_n(t)=\lambda(n+1)P_{n+1}(t)-\lambda n P_n(t)\;.
\label{eq mastersingleannih}
\end{equation}
This master equation can be solved exactly by different methods, see \cite{jahnke2007}. By means of the generating function \eqref{eq genfunc} Eq.~\eqref{eq mastersingleannih} can be written as ``Schr\"odinger equation''
\begin{equation}
i\lambda\frac{\partial}{\partial it}G(\xi,t)=\lambda^2(1-\xi)\frac{\partial}{\partial \xi}G(\xi,t)=i\lambda(1-\xi)\hat{\pi}G(\xi,t)\;,
\end{equation}
with the momentum operator $\hat{\pi}$. By applying the separation ansatz, we find the same formal solution for the time-dependent part as for the binary annihilation, see Eq.~\eqref{eq soltimebinary}. The $\xi$-dependent part of the separation ansatz is more interesting
\begin{equation}
E\varphi(\xi)=\lambda^2(1-\xi)\frac{\partial}{\partial \xi}\varphi(\xi)\;,
\end{equation}
it has the exact solution
\begin{equation}
\varphi(\xi)=\varphi(\xi_0)\left(\frac{\xi_0-1}{\xi-1}\right)^{E/\lambda^2}\;.
\label{eq solphiaingleann}
\end{equation}
The classical energy is $E=i\lambda(1-\xi)\pi$ and the classical equations of motion are
\begin{eqnarray}
\frac{d\xi}{d it}&=&i\lambda(1-\xi)\label{eq class1singleannih}\\
\frac{d\pi}{d it}&=&i\lambda\pi\;.
\end{eqnarray}
Both equations are easily solvable. The second equation gives the mean field dynamics resulting in the average particle number which is $\langle n\rangle=n_0e^{-\lambda t}$. The solution for \eqref{eq class1singleannih}
\begin{equation}
\xi=1+(\xi_0-1)e^{\lambda t}
\label{eq xisingleannhil}
\end{equation}
can be inserted into \eqref{eq solphiaingleann} which yields $\varphi(\xi)=\varphi(\xi_0)e^{-E t/\lambda}$. Hence, we find the generating function 
\begin{equation}
G(\xi,t)=\psi(t)\varphi(\xi)=\psi(0)\varphi(\xi_0)=G(\xi_0,0)=\xi^{n_0}_0\;.
\end{equation}
In the last step we used the initial condition that at time $t=0$ there are $n_0$ particles in the system. Solving Eq.~\eqref{eq xisingleannhil} for $\xi_0$ and inserting into the generating function yields
\begin{equation}
G(\xi,t)=\left(1+(\xi-1)e^{-\lambda t}\right)^{n_0}\;.
\end{equation}
The probability to find $n$ particles at time $t$ can be calculated by means of Eq.~\eqref{eq probderiv}. We find
\begin{equation}
	P_n(t)=\binom{n_0}{n}\left(1-e^{-\lambda t}\right)^{n_0-n}e^{-n\lambda t}\;.
\end{equation}

\subsection{Binary and single annihilation}\label{sec: bin and single}
\begin{figure}[ht]
 %\centering
 \includegraphics[width=0.48\textwidth]{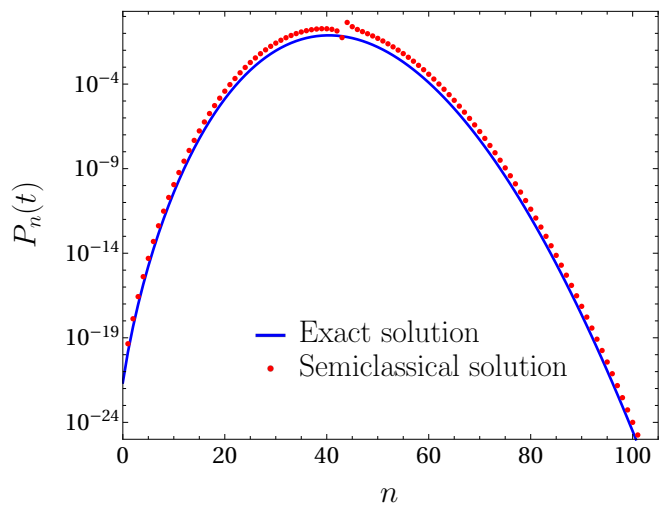}
  \caption{Probability density distribution for the binary and single annihilation, see Eq.~\eqref{eq masterbinsing}, at $t=2$ on a logarithmic scale. Parameters are chosen as $n_0=300$, $\lambda=0.01$ and $\sigma=0.05$.}
 \label{fig singlebinarycomp}
\end{figure}

We consider a system where two reactions can occur. The binary annihilation occurs with probability rate $\lambda$ and the single annihilation occurs with probability rate $\sigma$. Both probabilities are of the same order of magnitude. The master equation reads
\begin{eqnarray}
\frac{d}{dt}P_n(t)=&&\frac{\lambda}{2}\left((n+2)(n+1)P_{n+2}(t)-n(n-1)P_n(t)\right)\nonumber\\
&&+\sigma\left((n+1)P_{n+1}(t)-n P_n(t)\right)\;.
\label{eq masterbinsing}
\end{eqnarray}
Its corresponding ``Schr\"odinger equation'' for the generating function is
\begin{eqnarray}
i\lambda\frac{\partial}{\partial it}G(\xi,t)&&=\frac{\lambda^2}{2}(1-\xi^2)\frac{\partial^2}{\partial \xi^2}G(\xi,t)+\lambda\sigma(1-\xi)\frac{\partial}{\partial \xi}G(\xi,t)\nonumber\\
&&=-\frac{1}{2}(1-\xi^2)\hat{\pi}^2G(\xi,t)+i\sigma(1-\xi)\hat{\pi}G(\xi,t)\;,
\end{eqnarray}
where we use the momentum operator $\hat{\pi}=-i\lambda\partial/\partial\xi$ in the second line. The separation ansatz yields to the known result \eqref{eq soltimebinary} for the time-dependent part and for the stationary part we find
\begin{equation}
	E\varphi(\xi)=\frac{\lambda^2}{2}(1-\xi^2)\frac{\partial^2}{\partial \xi^2}\varphi(\xi)+\lambda\sigma(1-\xi)\frac{\partial}{\partial \xi}\varphi(\xi)\;.
\end{equation}
In the manner of section \ref{sec one} we use the ansatz $\varphi(\xi)=A(\xi)\exp(iS(\xi)/\lambda)$, sort the resulting equation for its real and imaginary part. Neglecting all terms with prefactor $\lambda^2$ or $\lambda\sigma$ we find
\begin{eqnarray}
E&&=\frac12 (\xi^2-1)\left(S'(\xi)\right)^2\label{eq wkbbinsingle1}\\
0&&=\frac{\lambda}{2}(\xi+1)\left(2A'(\xi)S'(\xi)+A(\xi)S''(\xi)\right)+\sigma A(\xi)S'(\xi)\;.\label{eq wkbbinsingle2}
\end{eqnarray}
The solution of \eqref{eq wkbbinsingle1} is \eqref{eq solSwkb} which is already discussed. We can insert the derivatives of $S(\xi)$ and insert them into \eqref{eq wkbbinsingle2} to obtain a differential equation for $A(\xi)$ with the solution
\begin{equation}
	A(\xi)=\tilde{c}(\xi^2-1)^{1/4}\left(\xi+1\right)^{-\sigma/\lambda}\;,
\end{equation}
with a constant $c$. Hence, the generating function, fulfilling the initial condition $G(\xi_0,0)=\xi^{n_0}_0$, is
\begin{eqnarray}
G(\xi,t)=&&\xi^{n_0}_0\left(\frac{\xi^2-1}{\xi^2_0-1}\right)^{1/4}\left(\frac{\xi_0+1}{\xi+1}\right)^{\sigma/\lambda}\nonumber\\
&&\times\exp\left(-\frac{\sqrt{2E}}{\lambda}(\arccos\xi-\arccos\xi_0)+\frac{1}{\lambda}Et\right)\;.
\end{eqnarray}
The classical energy is $E=\tfrac{1}{2}(\xi^2-1)\pi^2-i\sigma(\xi-1)\pi$ and the classical equations of motion are
\begin{eqnarray}
 \frac{d\xi}{d it}&&=(\xi^2-1)\pi-i\sigma(\xi-1)=\sqrt{2E(\xi^2-1)-\sigma^2(\xi-1)^2}\label{eq eqmotclasssinglebin}\\
 \frac{d\pi}{d it}&&=-\xi \pi^2+i\sigma\pi=-\pi\sqrt{2E+(\pi-i\sigma)^2}\;.
\end{eqnarray}
First, we evaluate the average particle number by means of the mean field dynamics
\begin{equation}
	\frac{d\bar{\pi}}{d it}=- \bar{\pi}^2+i\sigma\bar{\pi}\;.
\end{equation}
The solution gives the average particle number
\begin{equation}
	\langle n(t)\rangle=\frac{\sigma}{\lambda}\left(\frac{1}{1-\frac{\lambda n_0}{\lambda n_0+\sigma}e^{-\sigma t}}-1\right)\;.
\end{equation}
In order to make analytical progress we simplify Eq.~\eqref{eq eqmotclasssinglebin} furthermore by dropping the $\sigma^2$ term. This is fully consistent with the WKB method since by Taylor approximation the leading order of Eq.~\eqref{eq eqmotclasssinglebin} is $\sigma^2$. Due to this approximation we find the same classical dynamics as in the case of binary annihilation, see equations \eqref{eq class1} and \eqref{eq class2}. The solution for the corresponding classical equation of motion, determining $\xi$, is given in \eqref{eq solclass1}. Hence, the generating function has the same arguments in the exponential function as the generating function of the binary annihilation \eqref{eq genfunccE}. It solely differs from \eqref{eq genfunccE} by the additional prefactor $\left(\tfrac{\xi_0+1}{\xi+1}\right)^{\sigma/\lambda}$. We calculate the probability $P_n(t)$ by means of the contour integral with a saddle point approximation. The saddle point condition \eqref{eq solsaddle} combined with the solution of the approximated classical equation of motion \eqref{eq solclass1} gives the dependence
\begin{equation}
\xi_0=-\frac{\sigma}{\lambda n_0}+\sqrt{\left(1+\frac{\sigma}{\lambda n_0}\right)^2-\frac{n^2}{n^2_0}\frac{1-\xi^2_s}{\xi^2_s}}\;.
\end{equation}
Recombining this equation with \eqref{eq solclass1} yields an equation for $\xi_s$ which can be solved numerically. Finally the probability reads
\begin{eqnarray}
 P_n(t)=&&\frac{\xi^{n_0}_0}{\sqrt{2\pi}}\left(\frac{1+\xi_0}{1+\xi_s}\right)^{\sigma/\lambda}\left(\frac{1-\xi^2_s}{1-\xi^2_0}\right)^{1/4}\left|\frac{1-\xi^2_s}{n-\langle n\rangle\xi^2_s}\right|^{1/2}\nonumber\\
&&\times\exp\left(-\frac{n^2}{2}\lambda t\frac{1-\xi^2_s}{\xi^2_s}-n\ln\xi_s\right)\;.
\label{eq probsinglebinannihil}
\end{eqnarray}
We notice that the probability has the same form as the probability for the binary annihilation. The differences are the additional prefactor and the value for $\xi_s$ at the saddle point. The additional prefactor contributes only little to the value of \eqref{eq probsinglebinannihil} compared with the exponential. Furthermore, the value of the saddle point does not vary significantly from that of the binary annihilation. Hence, the probability $P_n(t)$ for the single and binary annihilation does not vary considerable from that for the pure binary annihilation. This means that the binary annihilation dominates the single annihilation considerably. In Fig.~\ref{fig singlebinarycomp} we compare the approximation with the exact probabilities.
We find some structure of the approximated probability near the maximum of the distribution. This is due to the prefactor which approaches zero in the denominator. Usually, in the saddle point approximation we consider the exponent only and do not consider the prefactor. If we do so, we would have to adapt an artificial prefactor in such a way that the normalization condition is still fulfilled. In that case the structure would vanish. Overall the approximation coincides over many orders of magnitude well with the exact solution.

\subsection{Third order reaction}\label{sec: third order}
\begin{figure}[t]
 %\centering
 \includegraphics[width=0.48\textwidth]{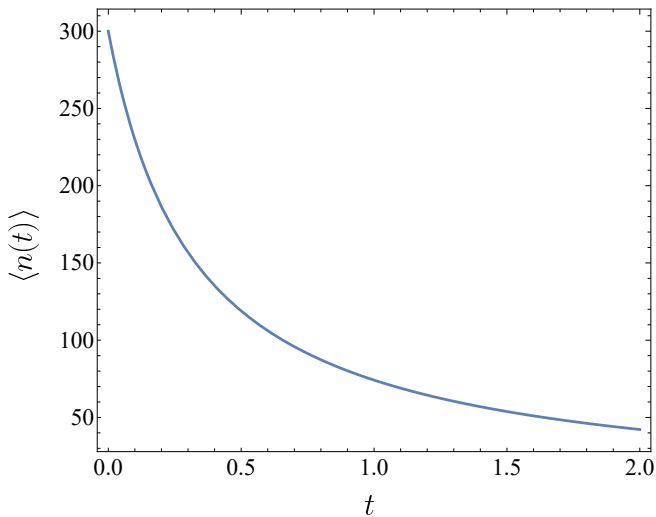}
  \caption{Dependence of the average particle number on the time $t$. Parameters are chosen as $n_0=300$, $\lambda=0.01$, $\sigma=\lambda$ and $\mu=\lambda^3$.}
 \label{fig avparttri}
\end{figure}

We study a reaction which combines the single and binary annihilation from the last example with the trimolecular annihilation. The master equation reads
\begin{eqnarray}
\frac{d}{dt} P_n(t)=&&\mu \dbinom{n+3}{3}P_{n+3}(t)+\lambda\dbinom{n+2}{2}P_{n+2}(t)\nonumber\\
&&+\sigma(n\!+\!1)P_{n\!+\!1}(t)\!-\!\left(\!\mu \dbinom{n}{3}\!+\!\lambda\dbinom{n}{2}\!+\!\sigma n\right)\!P_n(t)\;.
\label{eq mastertri}
\end{eqnarray}
We introduce the rate constant $\mu$ for the trimolecular reaction, which we assume to be of order $\lambda^3$. The rate constant $\sigma$ of the single annihilation is of order $\lambda$. The corresponding ``Schr\"odinger equation'' is
\begin{eqnarray}
i\lambda\frac{\partial}{\partial it}G(\xi,t)=&&\lambda\sigma(1-\xi)\frac{\partial}{\partial \xi}G(\xi,t)+\frac{\lambda^2}{2}(1-\xi^2)\frac{\partial^2}{\partial \xi^2}G(\xi,t)\nonumber\\
&&+\frac{\lambda \mu}{6}(1-\xi^3)\frac{\partial^3}{\partial \xi^3}G(\xi,t)\nonumber\\
=&&i\sigma(1-\xi)\hat{\pi}G(\xi,t)-\frac{1}{2}(1-\xi^2)\hat{\pi}^2G(\xi,t)\nonumber\\
&&-i\frac{\mu}{6\lambda^2}(1-\xi^3)\hat{\pi}^3G(\xi,t)\;,
\end{eqnarray}
with the momentum operator $\hat{\pi}=-i\lambda\partial/\partial\xi$. Applying the separation ansatz $G(\xi,t)=\psi(t)\varphi(\xi)$ yields to the result \eqref{eq soltimebinary} for the time dependent part. The stationary ``Schr\"odinger equation'' can be written as
\begin{eqnarray}
E\varphi(\xi)=&&\lambda\sigma(1-\xi)\frac{\partial}{\partial \xi}\varphi(\xi)+\frac{\lambda^2}{2}(1-\xi^2)\frac{\partial^2}{\partial \xi^2}\varphi(\xi)\nonumber\\
&&+\frac{\lambda \mu}{6}(1-\xi^3)\frac{\partial^3}{\partial \xi^3}\varphi(\xi)\;.
\end{eqnarray}
We insert the ansatz $\varphi(\xi)=A(\xi)\exp(iS(\xi)/\lambda)$ to solve this equation. We sort for the real and imaginary parts and drop all terms of order $\lambda^2$ and above which gives
\begin{eqnarray}
E&&=-\frac12 (1-\xi^2)\left(S'\right)^2\label{eq wkbtri1}\\
0&&=\sigma(1-\xi) AS'+\frac{\lambda}{2}(1-\xi^2)\left(2A'S'+AS''\right)\nonumber\\
&&-\frac{\mu}{6\lambda^2}(1-\xi^3)\left(S'\right)^3A\;.\label{eq wkbtri2}
\end{eqnarray}
For convenience of the reader, we dropped the arguments of $A(\xi)$ and $S(\xi)$. Once again we remark the scaling behavior $\sigma\sim\lambda$ and $\mu\sim\lambda^3$. Interestingly, we find Eq.~\eqref{eq wkbtri1} which we also found in the last example and also for the binary annihilation. Its solution is given in Eq.~\eqref{eq solSwkb}. Furthermore, each single term of Eq.~\eqref{eq wkbtri2} is of order $\lambda$. This means that we formally can cancel the small parameter $\lambda$ and so its order of magnitude does not appear in both equations. With $S'(\xi)$ from \eqref{eq wkbtri1} we can solve Eq.~\eqref{eq wkbtri2} and find
\begin{eqnarray}
A(\xi)=&&A_0(\xi_0)\exp\left(\frac{\mu E}{6\lambda^3}\frac{1}{1+\xi}\right)\left(\xi-1\right)^{(\lambda+\mu/\lambda^2E)/4\lambda}\nonumber\\
&&\times\left(\xi+1\right)^{(3\lambda-12\sigma+\mu/\lambda^2E)/12\lambda}\;.
\end{eqnarray}
With the initial condition $P_n(0)=\delta_{n,n_0}$ we are able to determine the constants. The probability now reads
\begin{eqnarray}
P_n(t)=&&\frac{1}{2\pi i}\xi^{n_0}_{0}\exp\left(-\frac{\mu E}{6\lambda^3}\frac{1}{1+\xi_0}\right)\nonumber\\
&&\times\oint{d\xi\,g(\xi,E)\exp\left(f(\xi,E)\right)}
\end{eqnarray}
with
\begin{eqnarray}
g(\xi,E)&&=\frac{1}{\xi}\left(\frac{\xi-1}{\xi_0-1}\right)^{\!\!(\lambda+\mu/\lambda^2E)/4\lambda}\!\!\nonumber\\
&&\times\left(\frac{\xi+1}{\xi_0+1}\right)^{\!\!(3\lambda-12\sigma+\mu/\lambda^2E)/12\lambda}\\
f(\xi,E)&&=\frac{\mu E}{6\lambda^3}\frac{1}{1+\xi}-\frac{\sqrt{2E}}{\lambda}(\arccos\xi-\arccos\xi_0)\nonumber\\
&&+\frac{Et}{\lambda}-n\ln\xi\;.
\end{eqnarray}
To make progress we specify the energy
\begin{equation}
E=i\sigma(1-\xi)\pi-\frac{1}{2}(1-\xi^2)\pi^2-i\frac{\mu}{6\lambda^2}(1-\xi^3)\pi^3
\label{eq entri}
\end{equation}
in terms of $\xi$. Therefore we have to solve the equations of motion
\begin{eqnarray}
\frac{d \xi}{d it}&&=i\sigma(1-\xi)-(1-\xi^2)\pi-i\frac{\mu}{2\lambda^2}(1-\xi^3)\pi^2\\ \label{eq eqmottri}
\frac{d \pi}{d it}&&=i\sigma\pi-\xi\pi^2-i\frac{\mu}{2\lambda^2}\xi^2\pi^3\;,
\end{eqnarray}
by inserting the solution of the energy for $\pi$ into Eq.~\eqref{eq eqmottri} which gives $\partial\xi/\partial it=\tau(\xi,E)$. The resulting equation cannot be solved analytically. Hence, we have to solve this equation by evaluating the integral \eqref{eq class1formal} numerically. However, this can only be done if we know the functional coherence of $\xi$ and $\xi_0$, which is obtained by the saddle point approximation, i.e., $f'(\xi_s,E)=0$. We find
\begin{eqnarray}
f'(\xi,E)=&&\frac{\mu E'}{6\lambda^3}\frac{1}{1+\xi}-\frac{\mu E}{6\lambda^3}\frac{1}{(1+\xi)^2}\nonumber\\
&&-\frac{\sqrt{2E'}}{\lambda}(\arccos\xi-\arccos\xi_0)\nonumber\\
&&+\frac{\sqrt{2E}}{\lambda}\frac{1}{\sqrt{1-\xi^2}}+\frac{t}{\lambda}E'-\frac{n}{\xi}
\end{eqnarray}
with $E'$ from Eq.~\eqref{eq eprime}. Again, this equation has to be solved numerically. A value for $\xi_s$ can be obtained by solving Eq.~\eqref{eq solinteqgeneral}. This procedure requires the numerical solution of two equations combined with a numerical integration. We remark that there are three possible solutions for $\pi$ when solving Eq.~\eqref{eq entri}. This is relevant for the numerics as we find $\xi_s<1$ for $n<\langle n\rangle$ and $\xi_s>1$ for $n>\langle n\rangle$ where $\langle n\rangle$ is the average particle number. By performing the numerics we have to use different solutions for $\pi$ when we calculate the values for $\xi_s$. This is depending on the value of $n$ being above or below $\langle n\rangle$. The mean field solution $\bar{\xi}=1$ gives us the dynamics of the average particle number
\begin{equation}
\frac{d \bar{\pi}}{d it}=i\sigma\bar{\pi}-\bar{\pi}^2-i\frac{\mu}{2\lambda^2}\bar{\pi}^3
\end{equation}
with $\bar{\pi}=-i\lambda\langle n\rangle$. We show the average particle number $\langle n(t)\rangle$ in Fig.~\ref{fig avparttri}. We specify the parameters $\lambda=\tfrac{1}{100}$, $\sigma=\lambda$, $\mu=\lambda^3$ and $n_0=300$. The average particle number at $t=1$ is $\langle n(1)\rangle\approx 74.15$. Hence, the transition $\xi_s>1$ arises when $n\geq75$ for $t=1$ if we only consider integer particle numbers.

\begin{figure}[ht]
 %\centering
 \includegraphics[width=0.48\textwidth]{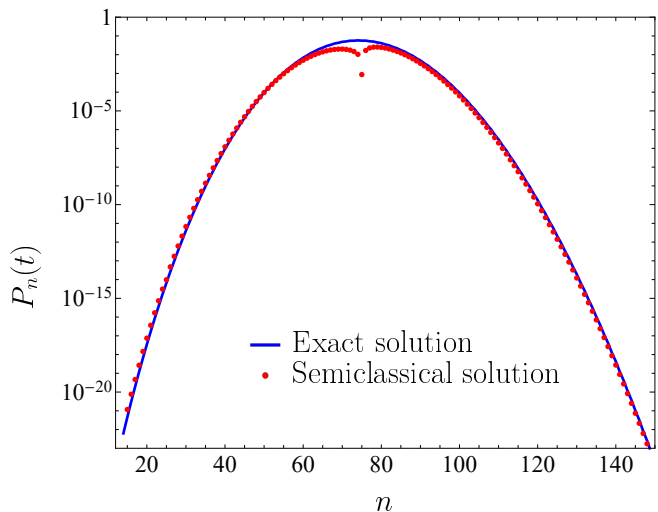}
  \caption{Probability density distribution for the combined annihilation up to third order, see Eq.~\eqref{eq mastertri}, at $t=1$ on a logarithmic scale. Parameters are chosen as $n_0=300$, $\lambda=0.01$, $\sigma=\lambda$ and $\mu=\lambda^3$.}
 \label{fig thirdorder}
\end{figure}

Having found the values for $\xi_s$ we can now perform the saddle point approximation which gives us
\begin{eqnarray}
P_n(t)=&&\frac{1}{2\pi}\xi^{n_0}_{0}\exp\left(-\frac{\mu E}{6\lambda^3}\frac{1}{1+\xi_0}\right)g(\xi_s,E_0)\nonumber\\
&&\times\dfrac{1}{\sqrt{\left|f''(\xi_s,E_0)\right|}}\exp\left(f(\xi_s,E_0)\right)
\end{eqnarray}
with the energy
\begin{equation}
E_0=\lambda\sigma(1-\xi_0)n_0+\frac{\lambda^2}{2}(1-\xi^2_0)n^2_0+\frac{\lambda \mu}{6}(1-\xi^3_0)n^3_0
\end{equation}
and the second derivative
\begin{widetext}
\begin{eqnarray}
f''(\xi,E)&&=E''\left(\frac{\mu}{6\lambda^3}\frac{1}{1+\xi}-\frac{1}{\sqrt{2E}\lambda}(\arccos\xi-\arccos\xi_0)+\frac{t}{\lambda}\right)\nonumber\\
&&+E'\left(-\frac{\mu}{3\lambda^3}\frac{1}{(1+\xi)^2}+\frac{E'}{(2E)^{3/2}\lambda}(\arccos\xi-\arccos\xi_0)+\frac{2}{\sqrt{2E}\lambda}\frac{1}{\sqrt{1-\xi^2}}\right)\nonumber\\
&&+\frac{\mu E}{3\lambda^3}\frac{1}{(1+\xi)^3}+\frac{\sqrt{2E}}{\lambda}\frac{\xi}{(1-\xi^2)^{3/2}}+\frac{n}{\xi^2}\;,
\end{eqnarray}
\end{widetext}
which is always real. In Fig.~\ref{fig thirdorder} we show the probability distribution $P_n(1)$ for the same parameter specification as in Fig.~\ref{fig avparttri}. Again we find a good agreement over many orders of magnitude between the semiclassical approximation and the exact solution, especially for the left tail of the distribution. 

\section{Conclusions}\label{sec conclude}
Extending the approach by Elgart and Kamenev, we put forward a model which is able to describe the probability of rare events in reaction-diffusion systems described by master equations. The systems consist of single-species particles with infinite-range interaction and we assume that spatial degrees of freedom are irrelevant. By means of a generating function we transform the master equation into a time-dependent ``Schr\"odinger equation'' in imaginary time. As the master equation gives a probabilistic description of the system, so does the evolution equation for the generating function. In short, the master equation is equivalent to the time-dependent ``Schr\"odinger equation'' derived by means of the generating function. We separate the time-dependent part of the ``Schr\"odinger equation'' and the stationary part and thereby introduce a new constant which we will interpret as the energy or Hamilton function of the system. We apply a WKB approximation to solve the stationary part of the ``Schr\"odinger equation'', we identify a small parameter which is the analogue to Planck's constant $\hbar$. In order to find the rare event statistics we are interested in large deviations from a typical system behavior. A typical behavior can be calculated by the mean field approximation which allows us to calculate quantities such as the average particle number. Within the WKB approximation we can derive the classical equations of motion which determine the phase portrait of the system. The trajectories of the phase portrait are determined by the energy which is constant in the semiclassical description. These classical equations of motion need to be solved either analytically or numerically in order to find the solution for the generating function. Finally, the probability to find $n$ particles at time $t$ is calculated by Cauchy's integral formula on which we apply a saddle point approximation. At this point we make a transition from the semiclassical description in which the dynamics is determined by the phase portrait to a probabilistic description. Hence, when performing the saddle point approximation the energy does not remain constant and becomes a function of the phase space parameters. This dependence is determined by the classical equations of motion. We remark that the accuracy of the method breaks down in the vicinity of the average particle number and for small times when the condition $\langle n(t)\rangle\ll n_0$ is not satisfied.

We studied some systems. The single annihilation is exactly solvable. The binary annihilation process and the process which combines binary and single annihilation is analytically solved by means of the WKB approximation. For both systems we find good agreement over many orders of magnitude with the exact solution of the master equation. Finally, we analyzed a process that combines single, binary and triple annihilation. In general, the leading order of the stationary ``Schr\"odinger equation'' is obtained by the order of the highest annihilation process. The third order differential equation can be solved analytically when we require a certain scaling of its parameters and use the WKB approximation. However, the semiclassical equations of motion have to be solved numerically. Once more, in the tails of the distribution, we find a very good agreement with the exact solution over many orders of magnitude.

Rare events define the tail of the distribution. We are not interested in typical fluctuations near the maximum of the distribution where we find oscillations for some examples. Nontheless, the oscillations can easily be removed by dropping the prefactor of the distribution and replacing it with a normalization constant. 

\begin{acknowledgments}
One of us (A.M.) acknowledges support from Studienstiftung des deutschen Volkes.
\end{acknowledgments}


\begin{thebibliography}{62}%
\makeatletter
\providecommand \@ifxundefined [1]{%
 \@ifx{#1\undefined}
}%
\providecommand \@ifnum [1]{%
 \ifnum #1\expandafter \@firstoftwo
 \else \expandafter \@secondoftwo
 \fi
}%
\providecommand \@ifx [1]{%
 \ifx #1\expandafter \@firstoftwo
 \else \expandafter \@secondoftwo
 \fi
}%
\providecommand \natexlab [1]{#1}%
\providecommand \enquote  [1]{``#1''}%
\providecommand \bibnamefont  [1]{#1}%
\providecommand \bibfnamefont [1]{#1}%
\providecommand \citenamefont [1]{#1}%
\providecommand \href@noop [0]{\@secondoftwo}%
\providecommand \href [0]{\begingroup \@sanitize@url \@href}%
\providecommand \@href[1]{\@@startlink{#1}\@@href}%
\providecommand \@@href[1]{\endgroup#1\@@endlink}%
\providecommand \@sanitize@url [0]{\catcode `\\12\catcode `\$12\catcode
  `\&12\catcode `\#12\catcode `\^12\catcode `\_12\catcode `\%12\relax}%
\providecommand \@@startlink[1]{}%
\providecommand \@@endlink[0]{}%
\providecommand \url  [0]{\begingroup\@sanitize@url \@url }%
\providecommand \@url [1]{\endgroup\@href {#1}{\urlprefix }}%
\providecommand \urlprefix  [0]{URL }%
\providecommand \Eprint [0]{\href }%
\providecommand \doibase [0]{http://dx.doi.org/}%
\providecommand \selectlanguage [0]{\@gobble}%
\providecommand \bibinfo  [0]{\@secondoftwo}%
\providecommand \bibfield  [0]{\@secondoftwo}%
\providecommand \translation [1]{[#1]}%
\providecommand \BibitemOpen [0]{}%
\providecommand \bibitemStop [0]{}%
\providecommand \bibitemNoStop [0]{.\EOS\space}%
\providecommand \EOS [0]{\spacefactor3000\relax}%
\providecommand \BibitemShut  [1]{\csname bibitem#1\endcsname}%
\let\auto@bib@innerbib\@empty
%</preamble>
\bibitem [{\citenamefont {M\"uhlbacher}\ and\ \citenamefont
  {Guhr}(2018{\natexlab{a}})}]{PubRev2018}%
  \BibitemOpen
  \bibfield  {author} {\bibinfo {author} {\bibfnamefont {A.}~\bibnamefont
  {M\"uhlbacher}}\ and\ \bibinfo {author} {\bibfnamefont {T.}~\bibnamefont
  {Guhr}},\ }\href@noop {} {\bibfield  {journal} {\bibinfo  {journal} {Risks}\
  }\textbf {\bibinfo {volume} {6}} (\bibinfo {year}
  {2018}{\natexlab{a}})}\BibitemShut {NoStop}%
\bibitem [{\citenamefont {Cardelli}\ \emph {et~al.}(2016)\citenamefont
  {Cardelli}, \citenamefont {Kwiatkowska},\ and\ \citenamefont
  {Laurenti}}]{cardelli2016}%
  \BibitemOpen
  \bibfield  {author} {\bibinfo {author} {\bibfnamefont {L.}~\bibnamefont
  {Cardelli}}, \bibinfo {author} {\bibfnamefont {M.}~\bibnamefont
  {Kwiatkowska}}, \ and\ \bibinfo {author} {\bibfnamefont {L.}~\bibnamefont
  {Laurenti}},\ }\href@noop {} {\bibfield  {journal} {\bibinfo  {journal}
  {Biosystems}\ }\textbf {\bibinfo {volume} {149}},\ \bibinfo {pages} {26}
  (\bibinfo {year} {2016})}\BibitemShut {NoStop}%
\bibitem [{\citenamefont {Van~Kampen}(2007)}]{van2007}%
  \BibitemOpen
  \bibfield  {author} {\bibinfo {author} {\bibfnamefont {N.~G.}\ \bibnamefont
  {Van~Kampen}},\ }\href@noop {} {\emph {\bibinfo {title} {Stochastic Processes
  in Physics and Chemistry}}},\ Vol.~\bibinfo {volume} {3}\ (\bibinfo
  {publisher} {Elsevier},\ \bibinfo {address} {Amsterdam},\ \bibinfo {year}
  {2007})\BibitemShut {NoStop}%
\bibitem [{\citenamefont {Bartlett}(1960)}]{bartlett1960}%
  \BibitemOpen
  \bibfield  {author} {\bibinfo {author} {\bibfnamefont {M.~S.}\ \bibnamefont
  {Bartlett}},\ }\href@noop {} {\emph {\bibinfo {title} {Stochastic Population
  Models in Ecology and Epidemiology}}}\ (\bibinfo  {publisher} {Wiley},\
  \bibinfo {address} {New York},\ \bibinfo {year} {1960})\BibitemShut {NoStop}%
\bibitem [{\citenamefont {Allen}(2010)}]{allen2010}%
  \BibitemOpen
  \bibfield  {author} {\bibinfo {author} {\bibfnamefont {L.~J.}\ \bibnamefont
  {Allen}},\ }\href@noop {} {\emph {\bibinfo {title} {An Introduction to
  Stochastic Processes with Applications to Biology}}}\ (\bibinfo  {publisher}
  {CRC Press},\ \bibinfo {address} {Hoboken, NJ},\ \bibinfo {year}
  {2010})\BibitemShut {NoStop}%
\bibitem [{\citenamefont {Nisbet}\ and\ \citenamefont
  {Gurney}(1982)}]{nisbet2003}%
  \BibitemOpen
  \bibfield  {author} {\bibinfo {author} {\bibfnamefont {R.~M.}\ \bibnamefont
  {Nisbet}}\ and\ \bibinfo {author} {\bibfnamefont {W.}~\bibnamefont
  {Gurney}},\ }\href@noop {} {\emph {\bibinfo {title} {Modelling Fluctuating
  Populations}}}\ (\bibinfo  {publisher} {Wiley},\ \bibinfo {address} {New
  York},\ \bibinfo {year} {1982})\BibitemShut {NoStop}%
\bibitem [{\citenamefont {N{\aa}sell}(2011)}]{naasell2011}%
  \BibitemOpen
  \bibfield  {author} {\bibinfo {author} {\bibfnamefont {I.}~\bibnamefont
  {N{\aa}sell}},\ }\href@noop {} {\emph {\bibinfo {title} {Extinction and
  Quasi-Stationarity in the Stochastic Logistic SIS Model}}}\ (\bibinfo
  {publisher} {Springer},\ \bibinfo {address} {Berlin},\ \bibinfo {year}
  {2011})\BibitemShut {NoStop}%
\bibitem [{\citenamefont {Altland}\ \emph {et~al.}(2011)\citenamefont
  {Altland}, \citenamefont {Fischer}, \citenamefont {Krug},\ and\ \citenamefont
  {Szendro}}]{krug2011}%
  \BibitemOpen
  \bibfield  {author} {\bibinfo {author} {\bibfnamefont {A.}~\bibnamefont
  {Altland}}, \bibinfo {author} {\bibfnamefont {A.}~\bibnamefont {Fischer}},
  \bibinfo {author} {\bibfnamefont {J.}~\bibnamefont {Krug}}, \ and\ \bibinfo
  {author} {\bibfnamefont {I.~G.}\ \bibnamefont {Szendro}},\ }\href {\doibase
  10.1103/PhysRevLett.106.088101} {\bibfield  {journal} {\bibinfo  {journal}
  {Phys. Rev. Lett.}\ }\textbf {\bibinfo {volume} {106}},\ \bibinfo {pages}
  {088101} (\bibinfo {year} {2011})}\BibitemShut {NoStop}%
\bibitem [{\citenamefont {Andersson}\ and\ \citenamefont
  {Britton}(2012)}]{andersson2012}%
  \BibitemOpen
  \bibfield  {author} {\bibinfo {author} {\bibfnamefont {H.}~\bibnamefont
  {Andersson}}\ and\ \bibinfo {author} {\bibfnamefont {T.}~\bibnamefont
  {Britton}},\ }\href@noop {} {\emph {\bibinfo {title} {Stochastic Epidemic
  Models and Their Statistical Analysis, Volume 151 of Lecture Notes in
  Statistics}}}\ (\bibinfo  {publisher} {Springer},\ \bibinfo {address} {New
  York, NY},\ \bibinfo {year} {2012})\BibitemShut {NoStop}%
\bibitem [{\citenamefont {Allen}\ \emph {et~al.}(2008)\citenamefont {Allen},
  \citenamefont {Brauer}, \citenamefont {Van~den Driessche},\ and\
  \citenamefont {Wu}}]{allen2008}%
  \BibitemOpen
  \bibfield  {author} {\bibinfo {author} {\bibfnamefont {L.~J.}\ \bibnamefont
  {Allen}}, \bibinfo {author} {\bibfnamefont {F.}~\bibnamefont {Brauer}},
  \bibinfo {author} {\bibfnamefont {P.}~\bibnamefont {Van~den Driessche}}, \
  and\ \bibinfo {author} {\bibfnamefont {J.}~\bibnamefont {Wu}},\ }\href@noop
  {} {\emph {\bibinfo {title} {Mathematical Epidemiology}}}\ (\bibinfo
  {publisher} {Springer},\ \bibinfo {address} {Berlin},\ \bibinfo {year}
  {2008})\BibitemShut {NoStop}%
\bibitem [{\citenamefont {Keeling}\ and\ \citenamefont
  {Ross}(2008)}]{keeling2008}%
  \BibitemOpen
  \bibfield  {author} {\bibinfo {author} {\bibfnamefont {M.~J.}\ \bibnamefont
  {Keeling}}\ and\ \bibinfo {author} {\bibfnamefont {J.~V.}\ \bibnamefont
  {Ross}},\ }\href@noop {} {\bibfield  {journal} {\bibinfo  {journal} {J. R.
  Soc. Interface}\ }\textbf {\bibinfo {volume} {5}},\ \bibinfo {pages} {171}
  (\bibinfo {year} {2008})}\BibitemShut {NoStop}%
\bibitem [{\citenamefont {Black}\ and\ \citenamefont
  {McKane}(2010)}]{black2010}%
  \BibitemOpen
  \bibfield  {author} {\bibinfo {author} {\bibfnamefont {A.~J.}\ \bibnamefont
  {Black}}\ and\ \bibinfo {author} {\bibfnamefont {A.~J.}\ \bibnamefont
  {McKane}},\ }\href@noop {} {\bibfield  {journal} {\bibinfo  {journal} {J. R.
  Soc. Interface}\ }\textbf {\bibinfo {volume} {7}},\ \bibinfo {pages} {1219}
  (\bibinfo {year} {2010})}\BibitemShut {NoStop}%
\bibitem [{\citenamefont {Allen}(2015)}]{allen2015}%
  \BibitemOpen
  \bibfield  {author} {\bibinfo {author} {\bibfnamefont {L.~J.}\ \bibnamefont
  {Allen}},\ }\href@noop {} {\emph {\bibinfo {title} {Stochastic Population and
  Epidemic Models}}},\ Vol.~\bibinfo {volume} {1}\ (\bibinfo  {publisher}
  {Springer International Publishing},\ \bibinfo {year} {2015})\BibitemShut
  {NoStop}%
\bibitem [{\citenamefont {M\"uhlbacher}\ and\ \citenamefont
  {Guhr}(2018{\natexlab{b}})}]{muehlbacher2018}%
  \BibitemOpen
  \bibfield  {author} {\bibinfo {author} {\bibfnamefont {A.}~\bibnamefont
  {M\"uhlbacher}}\ and\ \bibinfo {author} {\bibfnamefont {T.}~\bibnamefont
  {Guhr}},\ }\href@noop {} {\bibfield  {journal} {\bibinfo  {journal} {Risks}\
  }\textbf {\bibinfo {volume} {6}} (\bibinfo {year}
  {2018}{\natexlab{b}})}\BibitemShut {NoStop}%
\bibitem [{\citenamefont {Gardiner}(2009)}]{gardiner2009}%
  \BibitemOpen
  \bibfield  {author} {\bibinfo {author} {\bibfnamefont {C.}~\bibnamefont
  {Gardiner}},\ }\href@noop {} {\emph {\bibinfo {title} {Stochastic
  Methods}}},\ Vol.~\bibinfo {volume} {4}\ (\bibinfo  {publisher} {Springer},\
  \bibinfo {address} {Berlin},\ \bibinfo {year} {2009})\BibitemShut {NoStop}%
\bibitem [{\citenamefont {Mao}\ \emph {et~al.}(2002)\citenamefont {Mao},
  \citenamefont {Marion},\ and\ \citenamefont {Renshaw}}]{mao2002}%
  \BibitemOpen
  \bibfield  {author} {\bibinfo {author} {\bibfnamefont {X.}~\bibnamefont
  {Mao}}, \bibinfo {author} {\bibfnamefont {G.}~\bibnamefont {Marion}}, \ and\
  \bibinfo {author} {\bibfnamefont {E.}~\bibnamefont {Renshaw}},\ }\href@noop
  {} {\bibfield  {journal} {\bibinfo  {journal} {Stochastic Processes and Their
  Applications}\ }\textbf {\bibinfo {volume} {97}},\ \bibinfo {pages} {95}
  (\bibinfo {year} {2002})}\BibitemShut {NoStop}%
\bibitem [{\citenamefont {Roberts}\ \emph {et~al.}(2015)\citenamefont
  {Roberts}, \citenamefont {Be'er}, \citenamefont {Bohrer}, \citenamefont
  {Sharma},\ and\ \citenamefont {Assaf}}]{elijah2015}%
  \BibitemOpen
  \bibfield  {author} {\bibinfo {author} {\bibfnamefont {E.}~\bibnamefont
  {Roberts}}, \bibinfo {author} {\bibfnamefont {S.}~\bibnamefont {Be'er}},
  \bibinfo {author} {\bibfnamefont {C.}~\bibnamefont {Bohrer}}, \bibinfo
  {author} {\bibfnamefont {R.}~\bibnamefont {Sharma}}, \ and\ \bibinfo {author}
  {\bibfnamefont {M.}~\bibnamefont {Assaf}},\ }\href {\doibase
  10.1103/PhysRevE.92.062717} {\bibfield  {journal} {\bibinfo  {journal} {Phys.
  Rev. E}\ }\textbf {\bibinfo {volume} {92}},\ \bibinfo {pages} {062717}
  (\bibinfo {year} {2015})}\BibitemShut {NoStop}%
\bibitem [{\citenamefont {Lande}(1993)}]{lande1993}%
  \BibitemOpen
  \bibfield  {author} {\bibinfo {author} {\bibfnamefont {R.}~\bibnamefont
  {Lande}},\ }\href@noop {} {\bibfield  {journal} {\bibinfo  {journal} {The
  American Naturalist}\ }\textbf {\bibinfo {volume} {142}},\ \bibinfo {pages}
  {911} (\bibinfo {year} {1993})}\BibitemShut {NoStop}%
\bibitem [{\citenamefont {Kamenev}\ \emph {et~al.}(2008)\citenamefont
  {Kamenev}, \citenamefont {Meerson},\ and\ \citenamefont
  {Shklovskii}}]{kamenev2008}%
  \BibitemOpen
  \bibfield  {author} {\bibinfo {author} {\bibfnamefont {A.}~\bibnamefont
  {Kamenev}}, \bibinfo {author} {\bibfnamefont {B.}~\bibnamefont {Meerson}}, \
  and\ \bibinfo {author} {\bibfnamefont {B.}~\bibnamefont {Shklovskii}},\
  }\href {\doibase 10.1103/PhysRevLett.101.268103} {\bibfield  {journal}
  {\bibinfo  {journal} {Phys. Rev. Lett.}\ }\textbf {\bibinfo {volume} {101}},\
  \bibinfo {pages} {268103} (\bibinfo {year} {2008})}\BibitemShut {NoStop}%
\bibitem [{\citenamefont {Gillespie}(1992)}]{gillespie1992}%
  \BibitemOpen
  \bibfield  {author} {\bibinfo {author} {\bibfnamefont {D.~T.}\ \bibnamefont
  {Gillespie}},\ }\href@noop {} {\bibfield  {journal} {\bibinfo  {journal}
  {Physica A: Statistical Mechanics and its Applications}\ }\textbf {\bibinfo
  {volume} {188}},\ \bibinfo {pages} {404} (\bibinfo {year}
  {1992})}\BibitemShut {NoStop}%
\bibitem [{\citenamefont {Weber}\ and\ \citenamefont {Frey}(2017)}]{weber2017}%
  \BibitemOpen
  \bibfield  {author} {\bibinfo {author} {\bibfnamefont {M.~F.}\ \bibnamefont
  {Weber}}\ and\ \bibinfo {author} {\bibfnamefont {E.}~\bibnamefont {Frey}},\
  }\href@noop {} {\bibfield  {journal} {\bibinfo  {journal} {Rep. Prog. Phys.}\
  }\textbf {\bibinfo {volume} {80}},\ \bibinfo {pages} {046601} (\bibinfo
  {year} {2017})}\BibitemShut {NoStop}%
\bibitem [{\citenamefont {Jahnke}\ and\ \citenamefont
  {Huisinga}(2007)}]{jahnke2007}%
  \BibitemOpen
  \bibfield  {author} {\bibinfo {author} {\bibfnamefont {T.}~\bibnamefont
  {Jahnke}}\ and\ \bibinfo {author} {\bibfnamefont {W.}~\bibnamefont
  {Huisinga}},\ }\href@noop {} {\bibfield  {journal} {\bibinfo  {journal} {J.
  Math. Bio.}\ }\textbf {\bibinfo {volume} {54}},\ \bibinfo {pages} {1}
  (\bibinfo {year} {2007})}\BibitemShut {NoStop}%
\bibitem [{\citenamefont {McQuarrie}(1963)}]{mcquarrie1963}%
  \BibitemOpen
  \bibfield  {author} {\bibinfo {author} {\bibfnamefont {D.~A.}\ \bibnamefont
  {McQuarrie}},\ }\href {\doibase 10.1063/1.1733676} {\bibfield  {journal}
  {\bibinfo  {journal} {J. Chem. Phys.}\ }\textbf {\bibinfo {volume} {38}},\
  \bibinfo {pages} {433} (\bibinfo {year} {1963})}\BibitemShut {NoStop}%
\bibitem [{\citenamefont {McQuarrie}\ \emph {et~al.}(1964)\citenamefont
  {McQuarrie}, \citenamefont {Jachimowski},\ and\ \citenamefont
  {Russell}}]{mccquarrie1964}%
  \BibitemOpen
  \bibfield  {author} {\bibinfo {author} {\bibfnamefont {D.~A.}\ \bibnamefont
  {McQuarrie}}, \bibinfo {author} {\bibfnamefont {C.~J.}\ \bibnamefont
  {Jachimowski}}, \ and\ \bibinfo {author} {\bibfnamefont {M.~E.}\ \bibnamefont
  {Russell}},\ }\href {\doibase 10.1063/1.1724926} {\bibfield  {journal}
  {\bibinfo  {journal} {J. Chem. Phys.}\ }\textbf {\bibinfo {volume} {40}},\
  \bibinfo {pages} {2914} (\bibinfo {year} {1964})}\BibitemShut {NoStop}%
\bibitem [{\citenamefont {Laurenzi}(2000)}]{laurenzi2000}%
  \BibitemOpen
  \bibfield  {author} {\bibinfo {author} {\bibfnamefont {I.~J.}\ \bibnamefont
  {Laurenzi}},\ }\href {\doibase 10.1063/1.1287273} {\bibfield  {journal}
  {\bibinfo  {journal} {J. Chem. Phys.}\ }\textbf {\bibinfo {volume} {113}},\
  \bibinfo {pages} {3315} (\bibinfo {year} {2000})}\BibitemShut {NoStop}%
\bibitem [{\citenamefont {Walczak}\ \emph {et~al.}(2012)\citenamefont
  {Walczak}, \citenamefont {Mugler},\ and\ \citenamefont
  {Wiggins}}]{walczak2012}%
  \BibitemOpen
  \bibfield  {author} {\bibinfo {author} {\bibfnamefont {A.~M.}\ \bibnamefont
  {Walczak}}, \bibinfo {author} {\bibfnamefont {A.}~\bibnamefont {Mugler}}, \
  and\ \bibinfo {author} {\bibfnamefont {C.~H.}\ \bibnamefont {Wiggins}},\ }in\
  \href@noop {} {\emph {\bibinfo {booktitle} {Analytic Methods for Modeling
  Stochastic Regulatory Networks}}},\ \bibinfo {series and number} {Methods in
  Molecular Biology},\ \bibinfo {editor} {edited by\ \bibinfo {editor}
  {\bibfnamefont {X.}~\bibnamefont {Liu}}\ and\ \bibinfo {editor}
  {\bibfnamefont {M.}~\bibnamefont {Betterton}}}\ (\bibinfo  {publisher}
  {Humana Press, Totowa, NJ},\ \bibinfo {year} {2012})\ pp.\ \bibinfo {pages}
  {273--322}\BibitemShut {NoStop}%
\bibitem [{\citenamefont {Jenkinson}\ and\ \citenamefont
  {Goutsias}(2012)}]{jenkinson2012}%
  \BibitemOpen
  \bibfield  {author} {\bibinfo {author} {\bibfnamefont {G.}~\bibnamefont
  {Jenkinson}}\ and\ \bibinfo {author} {\bibfnamefont {J.}~\bibnamefont
  {Goutsias}},\ }\href@noop {} {\bibfield  {journal} {\bibinfo  {journal} {PLOS
  ONE}\ }\textbf {\bibinfo {volume} {7}} (\bibinfo {year} {2012})}\BibitemShut
  {NoStop}%
\bibitem [{\citenamefont {Schnoerr}\ \emph {et~al.}(2017)\citenamefont
  {Schnoerr}, \citenamefont {Sanguinetti},\ and\ \citenamefont
  {Grima}}]{schnoerr2017}%
  \BibitemOpen
  \bibfield  {author} {\bibinfo {author} {\bibfnamefont {D.}~\bibnamefont
  {Schnoerr}}, \bibinfo {author} {\bibfnamefont {G.}~\bibnamefont
  {Sanguinetti}}, \ and\ \bibinfo {author} {\bibfnamefont {R.}~\bibnamefont
  {Grima}},\ }\href@noop {} {\bibfield  {journal} {\bibinfo  {journal} {J.
  Phys. A: Math. Theor.}\ }\textbf {\bibinfo {volume} {50}},\ \bibinfo {pages}
  {093001} (\bibinfo {year} {2017})}\BibitemShut {NoStop}%
\bibitem [{\citenamefont {Li}\ \emph {et~al.}(2008)\citenamefont {Li},
  \citenamefont {Cao}, \citenamefont {Petzold},\ and\ \citenamefont
  {Gillespie}}]{li2008}%
  \BibitemOpen
  \bibfield  {author} {\bibinfo {author} {\bibfnamefont {H.}~\bibnamefont
  {Li}}, \bibinfo {author} {\bibfnamefont {Y.}~\bibnamefont {Cao}}, \bibinfo
  {author} {\bibfnamefont {L.~R.}\ \bibnamefont {Petzold}}, \ and\ \bibinfo
  {author} {\bibfnamefont {D.~T.}\ \bibnamefont {Gillespie}},\ }\href {\doibase
  10.1021/bp070255h} {\bibfield  {journal} {\bibinfo  {journal} {Biotechnology
  Progress}\ }\textbf {\bibinfo {volume} {24}},\ \bibinfo {pages} {56}
  (\bibinfo {year} {2008})}\BibitemShut {NoStop}%
\bibitem [{\citenamefont {Pahle}(2009)}]{pahle2009}%
  \BibitemOpen
  \bibfield  {author} {\bibinfo {author} {\bibfnamefont {J.}~\bibnamefont
  {Pahle}},\ }\href {\doibase 10.1093/bib/bbn050} {\bibfield  {journal}
  {\bibinfo  {journal} {Briefings in Bioinformatics}\ }\textbf {\bibinfo
  {volume} {10}},\ \bibinfo {pages} {53} (\bibinfo {year} {2009})}\BibitemShut
  {NoStop}%
\bibitem [{\citenamefont {Gillespie}\ \emph {et~al.}(2013)\citenamefont
  {Gillespie}, \citenamefont {Hellander},\ and\ \citenamefont
  {Petzold}}]{gillespie2013}%
  \BibitemOpen
  \bibfield  {author} {\bibinfo {author} {\bibfnamefont {D.~T.}\ \bibnamefont
  {Gillespie}}, \bibinfo {author} {\bibfnamefont {A.}~\bibnamefont
  {Hellander}}, \ and\ \bibinfo {author} {\bibfnamefont {L.~R.}\ \bibnamefont
  {Petzold}},\ }\href {\doibase 10.1063/1.4801941} {\bibfield  {journal}
  {\bibinfo  {journal} {J. Chem. Phys.}\ }\textbf {\bibinfo {volume} {138}},\
  \bibinfo {pages} {170901} (\bibinfo {year} {2013})}\BibitemShut {NoStop}%
\bibitem [{\citenamefont {Mauch}\ and\ \citenamefont
  {Stalzer}(2011)}]{Mauch2011}%
  \BibitemOpen
  \bibfield  {author} {\bibinfo {author} {\bibfnamefont {S.}~\bibnamefont
  {Mauch}}\ and\ \bibinfo {author} {\bibfnamefont {M.}~\bibnamefont
  {Stalzer}},\ }\href {\doibase 10.1109/TCBB.2009.47} {\bibfield  {journal}
  {\bibinfo  {journal} {IEEE/ACM Trans. on Computational Biology and
  Bioinformatics}\ }\textbf {\bibinfo {volume} {8}},\ \bibinfo {pages} {27}
  (\bibinfo {year} {2011})}\BibitemShut {NoStop}%
\bibitem [{\citenamefont {Gillespie}(1976)}]{gillespie1976}%
  \BibitemOpen
  \bibfield  {author} {\bibinfo {author} {\bibfnamefont {D.~T.}\ \bibnamefont
  {Gillespie}},\ }\href@noop {} {\bibfield  {journal} {\bibinfo  {journal} {J.
  Comput. Phys.}\ }\textbf {\bibinfo {volume} {22}},\ \bibinfo {pages} {403}
  (\bibinfo {year} {1976})}\BibitemShut {NoStop}%
\bibitem [{\citenamefont {Gillespie}(1977)}]{gillespie1977}%
  \BibitemOpen
  \bibfield  {author} {\bibinfo {author} {\bibfnamefont {D.~T.}\ \bibnamefont
  {Gillespie}},\ }\href {\doibase 10.1021/j100540a008} {\bibfield  {journal}
  {\bibinfo  {journal} {J. Phys. Chem.}\ }\textbf {\bibinfo {volume} {81}},\
  \bibinfo {pages} {2340} (\bibinfo {year} {1977})}\BibitemShut {NoStop}%
\bibitem [{\citenamefont {Gillespie}(2007)}]{gillespie2007}%
  \BibitemOpen
  \bibfield  {author} {\bibinfo {author} {\bibfnamefont {D.~T.}\ \bibnamefont
  {Gillespie}},\ }\href {\doibase 10.1146/annurev.physchem.58.032806.104637}
  {\bibfield  {journal} {\bibinfo  {journal} {Annu. Rev. Phys. Chem.}\ }\textbf
  {\bibinfo {volume} {58}},\ \bibinfo {pages} {35} (\bibinfo {year}
  {2007})}\BibitemShut {NoStop}%
\bibitem [{\citenamefont {Risken}(1996)}]{risken1996}%
  \BibitemOpen
  \bibfield  {author} {\bibinfo {author} {\bibfnamefont {H.}~\bibnamefont
  {Risken}},\ }\href@noop {} {\emph {\bibinfo {title} {The Fokker-Planck
  Equation}}},\ Vol.~\bibinfo {volume} {2}\ (\bibinfo  {publisher} {Springer},\
  \bibinfo {address} {Berlin},\ \bibinfo {year} {1996})\BibitemShut {NoStop}%
\bibitem [{\citenamefont {Doering}\ \emph {et~al.}(2005)\citenamefont
  {Doering}, \citenamefont {Sargsyan},\ and\ \citenamefont
  {Sander}}]{doering2005}%
  \BibitemOpen
  \bibfield  {author} {\bibinfo {author} {\bibfnamefont {C.}~\bibnamefont
  {Doering}}, \bibinfo {author} {\bibfnamefont {K.}~\bibnamefont {Sargsyan}}, \
  and\ \bibinfo {author} {\bibfnamefont {L.}~\bibnamefont {Sander}},\ }\href
  {\doibase 10.1137/030602800} {\bibfield  {journal} {\bibinfo  {journal}
  {Multiscale Modeling \& Simulation}\ }\textbf {\bibinfo {volume} {3}},\
  \bibinfo {pages} {283} (\bibinfo {year} {2005})}\BibitemShut {NoStop}%
\bibitem [{\citenamefont {Gaveau}\ \emph {et~al.}(1996)\citenamefont {Gaveau},
  \citenamefont {Moreau},\ and\ \citenamefont {Toth}}]{gaveau1996}%
  \BibitemOpen
  \bibfield  {author} {\bibinfo {author} {\bibfnamefont {B.}~\bibnamefont
  {Gaveau}}, \bibinfo {author} {\bibfnamefont {M.}~\bibnamefont {Moreau}}, \
  and\ \bibinfo {author} {\bibfnamefont {J.}~\bibnamefont {Toth}},\ }\href@noop
  {} {\bibfield  {journal} {\bibinfo  {journal} {Letters in Mathematical
  Physics}\ }\textbf {\bibinfo {volume} {37}},\ \bibinfo {pages} {285}
  (\bibinfo {year} {1996})}\BibitemShut {NoStop}%
\bibitem [{\citenamefont {Elgart}\ and\ \citenamefont
  {Kamenev}(2004)}]{ElgartKamenev2004}%
  \BibitemOpen
  \bibfield  {author} {\bibinfo {author} {\bibfnamefont {V.}~\bibnamefont
  {Elgart}}\ and\ \bibinfo {author} {\bibfnamefont {A.}~\bibnamefont
  {Kamenev}},\ }\href {\doibase 10.1103/PhysRevE.70.041106} {\bibfield
  {journal} {\bibinfo  {journal} {Phys. Rev. E}\ }\textbf {\bibinfo {volume}
  {70}},\ \bibinfo {pages} {041106} (\bibinfo {year} {2004})}\BibitemShut
  {NoStop}%
\bibitem [{\citenamefont {Landau}\ and\ \citenamefont
  {Lifshitz}(1977)}]{landau1977}%
  \BibitemOpen
  \bibfield  {author} {\bibinfo {author} {\bibfnamefont {L.~D.}\ \bibnamefont
  {Landau}}\ and\ \bibinfo {author} {\bibfnamefont {E.~M.}\ \bibnamefont
  {Lifshitz}},\ }\href@noop {} {\emph {\bibinfo {title} {Quantum Mechanics}}},\
  Vol.~\bibinfo {volume} {3}\ (\bibinfo  {publisher} {Pergamon},\ \bibinfo
  {address} {Oxford},\ \bibinfo {year} {1977})\BibitemShut {NoStop}%
\bibitem [{\citenamefont {Miller}\ and\ \citenamefont
  {Good}(1953)}]{miller1953}%
  \BibitemOpen
  \bibfield  {author} {\bibinfo {author} {\bibfnamefont {S.~C.}\ \bibnamefont
  {Miller}}\ and\ \bibinfo {author} {\bibfnamefont {R.~H.}\ \bibnamefont
  {Good}},\ }\href {\doibase 10.1103/PhysRev.91.174} {\bibfield  {journal}
  {\bibinfo  {journal} {Phys. Rev.}\ }\textbf {\bibinfo {volume} {91}},\
  \bibinfo {pages} {174} (\bibinfo {year} {1953})}\BibitemShut {NoStop}%
\bibitem [{\citenamefont {Messiah}(1967)}]{messiah1964}%
  \BibitemOpen
  \bibfield  {author} {\bibinfo {author} {\bibfnamefont {A.}~\bibnamefont
  {Messiah}},\ }\href@noop {} {\emph {\bibinfo {title} {Quantum Mechanics}}}\
  (\bibinfo  {publisher} {North-Holland Publ.},\ \bibinfo {address}
  {Amsterdam},\ \bibinfo {year} {1967})\BibitemShut {NoStop}%
\bibitem [{\citenamefont {Griffiths}\ and\ \citenamefont
  {Schroeter}(2018)}]{griffiths2018}%
  \BibitemOpen
  \bibfield  {author} {\bibinfo {author} {\bibfnamefont {D.~J.}\ \bibnamefont
  {Griffiths}}\ and\ \bibinfo {author} {\bibfnamefont {D.~F.}\ \bibnamefont
  {Schroeter}},\ }\href {\doibase 10.1017/9781316995433} {\emph {\bibinfo
  {title} {Introduction to Quantum Mechanics}}},\ \bibinfo {edition} {3rd}\
  ed.\ (\bibinfo  {publisher} {Cambridge University Press},\ \bibinfo {address}
  {Cambridge},\ \bibinfo {year} {2018})\BibitemShut {NoStop}%
\bibitem [{\citenamefont {Bender}\ and\ \citenamefont
  {Orszag}(2013)}]{bender2013}%
  \BibitemOpen
  \bibfield  {author} {\bibinfo {author} {\bibfnamefont {C.~M.}\ \bibnamefont
  {Bender}}\ and\ \bibinfo {author} {\bibfnamefont {S.~A.}\ \bibnamefont
  {Orszag}},\ }\href@noop {} {\emph {\bibinfo {title} {Advanced Mathematical
  Methods for Scientists and Engineers I: Asymptotic Methods and Perturbation
  Theory}}}\ (\bibinfo  {publisher} {Springer Science \& Business Media},\
  \bibinfo {address} {New York},\ \bibinfo {year} {2013})\BibitemShut {NoStop}%
\bibitem [{\citenamefont {Assaf}\ and\ \citenamefont
  {Meerson}(2006)}]{assaf2006}%
  \BibitemOpen
  \bibfield  {author} {\bibinfo {author} {\bibfnamefont {M.}~\bibnamefont
  {Assaf}}\ and\ \bibinfo {author} {\bibfnamefont {B.}~\bibnamefont
  {Meerson}},\ }\href {\doibase 10.1103/PhysRevE.74.041115} {\bibfield
  {journal} {\bibinfo  {journal} {Phys. Rev. E}\ }\textbf {\bibinfo {volume}
  {74}},\ \bibinfo {pages} {041115} (\bibinfo {year} {2006})}\BibitemShut
  {NoStop}%
\bibitem [{\citenamefont {Assaf}\ \emph {et~al.}(2010)\citenamefont {Assaf},
  \citenamefont {Meerson},\ and\ \citenamefont {Sasorov}}]{assaf2010}%
  \BibitemOpen
  \bibfield  {author} {\bibinfo {author} {\bibfnamefont {M.}~\bibnamefont
  {Assaf}}, \bibinfo {author} {\bibfnamefont {B.}~\bibnamefont {Meerson}}, \
  and\ \bibinfo {author} {\bibfnamefont {P.~V.}\ \bibnamefont {Sasorov}},\
  }\href@noop {} {\bibfield  {journal} {\bibinfo  {journal} {J. Stat. Mech.}\
  }\textbf {\bibinfo {volume} {2010}},\ \bibinfo {pages} {P07018} (\bibinfo
  {year} {2010})}\BibitemShut {NoStop}%
\bibitem [{\citenamefont {Assaf}\ and\ \citenamefont
  {Meerson}(2017{\natexlab{a}})}]{assaf2017}%
  \BibitemOpen
  \bibfield  {author} {\bibinfo {author} {\bibfnamefont {M.}~\bibnamefont
  {Assaf}}\ and\ \bibinfo {author} {\bibfnamefont {B.}~\bibnamefont
  {Meerson}},\ }\href@noop {} {\bibfield  {journal} {\bibinfo  {journal} {J.
  Phys. A}\ }\textbf {\bibinfo {volume} {50}},\ \bibinfo {pages} {263001}
  (\bibinfo {year} {2017}{\natexlab{a}})}\BibitemShut {NoStop}%
\bibitem [{\citenamefont {Be'er}\ \emph {et~al.}(2015)\citenamefont {Be'er},
  \citenamefont {Assaf},\ and\ \citenamefont {Meerson}}]{beer2015}%
  \BibitemOpen
  \bibfield  {author} {\bibinfo {author} {\bibfnamefont {S.}~\bibnamefont
  {Be'er}}, \bibinfo {author} {\bibfnamefont {M.}~\bibnamefont {Assaf}}, \ and\
  \bibinfo {author} {\bibfnamefont {B.}~\bibnamefont {Meerson}},\ }\href
  {\doibase 10.1103/PhysRevE.91.062126} {\bibfield  {journal} {\bibinfo
  {journal} {Phys. Rev. E}\ }\textbf {\bibinfo {volume} {91}},\ \bibinfo
  {pages} {062126} (\bibinfo {year} {2015})}\BibitemShut {NoStop}%
\bibitem [{\citenamefont {Be’er}\ and\ \citenamefont
  {Assaf}(2016)}]{beer2016}%
  \BibitemOpen
  \bibfield  {author} {\bibinfo {author} {\bibfnamefont {S.}~\bibnamefont
  {Be’er}}\ and\ \bibinfo {author} {\bibfnamefont {M.}~\bibnamefont
  {Assaf}},\ }\href@noop {} {\bibfield  {journal} {\bibinfo  {journal} {J.
  Stat. Mech.}\ }\textbf {\bibinfo {volume} {2016}},\ \bibinfo {pages} {113501}
  (\bibinfo {year} {2016})}\BibitemShut {NoStop}%
\bibitem [{\citenamefont {Dykman}\ \emph {et~al.}(1994)\citenamefont {Dykman},
  \citenamefont {Mori}, \citenamefont {Ross},\ and\ \citenamefont
  {Hunt}}]{dykman1994}%
  \BibitemOpen
  \bibfield  {author} {\bibinfo {author} {\bibfnamefont {M.~I.}\ \bibnamefont
  {Dykman}}, \bibinfo {author} {\bibfnamefont {E.}~\bibnamefont {Mori}},
  \bibinfo {author} {\bibfnamefont {J.}~\bibnamefont {Ross}}, \ and\ \bibinfo
  {author} {\bibfnamefont {P.~M.}\ \bibnamefont {Hunt}},\ }\href {\doibase
  10.1063/1.467139} {\bibfield  {journal} {\bibinfo  {journal} {The Journal of
  Chemical Physics}\ }\textbf {\bibinfo {volume} {100}},\ \bibinfo {pages}
  {5735} (\bibinfo {year} {1994})}\BibitemShut {NoStop}%
\bibitem [{\citenamefont {Gang}(1987)}]{gang1987}%
  \BibitemOpen
  \bibfield  {author} {\bibinfo {author} {\bibfnamefont {H.}~\bibnamefont
  {Gang}},\ }\href {\doibase 10.1103/PhysRevA.36.5782} {\bibfield  {journal}
  {\bibinfo  {journal} {Phys. Rev. A}\ }\textbf {\bibinfo {volume} {36}},\
  \bibinfo {pages} {5782} (\bibinfo {year} {1987})}\BibitemShut {NoStop}%
\bibitem [{\citenamefont {Kubo}\ \emph {et~al.}(1973)\citenamefont {Kubo},
  \citenamefont {Matsuo},\ and\ \citenamefont {Kitahara}}]{kubo1973}%
  \BibitemOpen
  \bibfield  {author} {\bibinfo {author} {\bibfnamefont {R.}~\bibnamefont
  {Kubo}}, \bibinfo {author} {\bibfnamefont {K.}~\bibnamefont {Matsuo}}, \ and\
  \bibinfo {author} {\bibfnamefont {K.}~\bibnamefont {Kitahara}},\ }\href
  {\doibase 10.1007/BF01016797} {\bibfield  {journal} {\bibinfo  {journal}
  {Journal of Statistical Physics}\ }\textbf {\bibinfo {volume} {9}},\ \bibinfo
  {pages} {51} (\bibinfo {year} {1973})}\BibitemShut {NoStop}%
\bibitem [{\citenamefont {Peters}\ \emph {et~al.}(1989)\citenamefont {Peters},
  \citenamefont {Mangel},\ and\ \citenamefont {Costantino}}]{peters1989}%
  \BibitemOpen
  \bibfield  {author} {\bibinfo {author} {\bibfnamefont {C.~S.}\ \bibnamefont
  {Peters}}, \bibinfo {author} {\bibfnamefont {M.}~\bibnamefont {Mangel}}, \
  and\ \bibinfo {author} {\bibfnamefont {R.}~\bibnamefont {Costantino}},\
  }\href@noop {} {\bibfield  {journal} {\bibinfo  {journal} {Bull. Math. Bio.}\
  }\textbf {\bibinfo {volume} {51}},\ \bibinfo {pages} {625} (\bibinfo {year}
  {1989})}\BibitemShut {NoStop}%
\bibitem [{\citenamefont {Assaf}\ and\ \citenamefont
  {Meerson}(2017{\natexlab{b}})}]{assaf2017rev}%
  \BibitemOpen
  \bibfield  {author} {\bibinfo {author} {\bibfnamefont {M.}~\bibnamefont
  {Assaf}}\ and\ \bibinfo {author} {\bibfnamefont {B.}~\bibnamefont
  {Meerson}},\ }\href@noop {} {\bibfield  {journal} {\bibinfo  {journal} {J.
  Phys. A}\ }\textbf {\bibinfo {volume} {50}},\ \bibinfo {pages} {263001}
  (\bibinfo {year} {2017}{\natexlab{b}})}\BibitemShut {NoStop}%
\bibitem [{\citenamefont {Ovaskainen}\ and\ \citenamefont
  {Meerson}(2010)}]{ovaskainen2010}%
  \BibitemOpen
  \bibfield  {author} {\bibinfo {author} {\bibfnamefont {O.}~\bibnamefont
  {Ovaskainen}}\ and\ \bibinfo {author} {\bibfnamefont {B.}~\bibnamefont
  {Meerson}},\ }\href@noop {} {\bibfield  {journal} {\bibinfo  {journal}
  {Trends in Ecology \& Evolution}\ }\textbf {\bibinfo {volume} {25}},\
  \bibinfo {pages} {643} (\bibinfo {year} {2010})}\BibitemShut {NoStop}%
\bibitem [{\citenamefont {Bressloff}(2017)}]{bressloff2017}%
  \BibitemOpen
  \bibfield  {author} {\bibinfo {author} {\bibfnamefont {P.~C.}\ \bibnamefont
  {Bressloff}},\ }\href@noop {} {\bibfield  {journal} {\bibinfo  {journal} {J.
  Phys. A}\ }\textbf {\bibinfo {volume} {50}},\ \bibinfo {pages} {133001}
  (\bibinfo {year} {2017})}\BibitemShut {NoStop}%
\bibitem [{\citenamefont {Assaf}\ and\ \citenamefont
  {Meerson}(2007)}]{assaf2007ex}%
  \BibitemOpen
  \bibfield  {author} {\bibinfo {author} {\bibfnamefont {M.}~\bibnamefont
  {Assaf}}\ and\ \bibinfo {author} {\bibfnamefont {B.}~\bibnamefont
  {Meerson}},\ }\href {\doibase 10.1103/PhysRevE.75.031122} {\bibfield
  {journal} {\bibinfo  {journal} {Phys. Rev. E}\ }\textbf {\bibinfo {volume}
  {75}},\ \bibinfo {pages} {031122} (\bibinfo {year} {2007})}\BibitemShut
  {NoStop}%
\bibitem [{\citenamefont {Assaf}\ and\ \citenamefont
  {Meerson}(2010)}]{assaf2010ext}%
  \BibitemOpen
  \bibfield  {author} {\bibinfo {author} {\bibfnamefont {M.}~\bibnamefont
  {Assaf}}\ and\ \bibinfo {author} {\bibfnamefont {B.}~\bibnamefont
  {Meerson}},\ }\href {\doibase 10.1103/PhysRevE.81.021116} {\bibfield
  {journal} {\bibinfo  {journal} {Phys. Rev. E}\ }\textbf {\bibinfo {volume}
  {81}},\ \bibinfo {pages} {021116} (\bibinfo {year} {2010})}\BibitemShut
  {NoStop}%
\bibitem [{\citenamefont {Kessler}\ and\ \citenamefont
  {Shnerb}(2007)}]{kessler2007}%
  \BibitemOpen
  \bibfield  {author} {\bibinfo {author} {\bibfnamefont {D.~A.}\ \bibnamefont
  {Kessler}}\ and\ \bibinfo {author} {\bibfnamefont {N.~M.}\ \bibnamefont
  {Shnerb}},\ }\href@noop {} {\bibfield  {journal} {\bibinfo  {journal} {J.
  Stat. Phys.}\ }\textbf {\bibinfo {volume} {127}},\ \bibinfo {pages} {861}
  (\bibinfo {year} {2007})}\BibitemShut {NoStop}%
\bibitem [{\citenamefont {Assaf}\ \emph {et~al.}(2009)\citenamefont {Assaf},
  \citenamefont {Kamenev},\ and\ \citenamefont {Meerson}}]{assaf2009ex}%
  \BibitemOpen
  \bibfield  {author} {\bibinfo {author} {\bibfnamefont {M.}~\bibnamefont
  {Assaf}}, \bibinfo {author} {\bibfnamefont {A.}~\bibnamefont {Kamenev}}, \
  and\ \bibinfo {author} {\bibfnamefont {B.}~\bibnamefont {Meerson}},\ }\href
  {\doibase 10.1103/PhysRevE.79.011127} {\bibfield  {journal} {\bibinfo
  {journal} {Phys. Rev. E}\ }\textbf {\bibinfo {volume} {79}},\ \bibinfo
  {pages} {011127} (\bibinfo {year} {2009})}\BibitemShut {NoStop}%
\bibitem [{\citenamefont {Escudero}\ and\ \citenamefont
  {Kamenev}(2009)}]{escudero2009}%
  \BibitemOpen
  \bibfield  {author} {\bibinfo {author} {\bibfnamefont {C.}~\bibnamefont
  {Escudero}}\ and\ \bibinfo {author} {\bibfnamefont {A.}~\bibnamefont
  {Kamenev}},\ }\href {\doibase 10.1103/PhysRevE.79.041149} {\bibfield
  {journal} {\bibinfo  {journal} {Phys. Rev. E}\ }\textbf {\bibinfo {volume}
  {79}},\ \bibinfo {pages} {041149} (\bibinfo {year} {2009})}\BibitemShut
  {NoStop}%
\bibitem [{\citenamefont {Chang}(1968)}]{chang1968}%
  \BibitemOpen
  \bibfield  {author} {\bibinfo {author} {\bibfnamefont {C.~L.}\ \bibnamefont
  {Chang}},\ }\href@noop {} {\emph {\bibinfo {title} {Introduction to
  Stochastic Processes in Biostatistics}}}\ (\bibinfo  {publisher} {John Wiley
  And Sons, Inc; New York},\ \bibinfo {year} {1968})\BibitemShut {NoStop}%
\end{thebibliography}
\end{document}